# Designable Dynamical Systems for the Generalized Landau Scenario and the Nonlinear Complexification of Periodic Orbits

R. Herrero[*], J. Farjas[†], F. Pi[‡], G. Orriols[‡]

[*] Departament de Física i Enginyeria Nuclear,
Universitat Politècnica de Catalunya,
08222 Terrassa, Spain
ramon.herrero@upc.edu
[†] Departament de Física, Campus Montilivi,
Universitat de Girona,
17003 Girona, Spain
jordi.farjas@udg.edu
[‡] Departament de Física, Universitat Autònoma de Barcelona,
08193 Cerdanyola del Vallès, Spain
francesc.pi@uab.cat
gaspar.orriols@uab.cat

We have found a way for penetrating the space of the dynamical systems towards systems of arbitrary dimension exhibiting the nonlinear mixing of a significant number of oscillation modes through which extraordinarily complex time evolutions arise. The system design is based on assuring the occurrence of a number of Hopf bifurcations in a set of fixed points of a relatively generic system of ordinary differential equations, in which the main peculiarity is that the nonlinearities appear through functions of a linear combination of the system variables. The paper presents the design procedure and a selection of numerical simulations with a variety of designed systems whose dynamical behaviours are really rich and full of unknown features. For concreteness, we focus the presentation to illustrating the oscillatory mixing effects on the periodic orbits, through which the harmonic oscillation born in a Hopf bifurcation becomes successively enriched with the intermittent incorporation of other oscillation modes of higher frequencies while the orbit remains periodic and without necessity of bifurcating instabilities. Even in the absence of a proper mathematical theory covering the nonlinear mixing mechanisms we find enough evidence to expect that the oscillatory scenario be truly scalable concerning the phase space dimension, the multiplicity of involved fixed points and the range of time scales, so that extremely complex but ordered dynamical behaviours could be sustained through it.

## 1. Introduction

The typical approach in applied nonlinear dynamics is modelling: The tentative achievement of a proper set of differential equations whose dynamical behaviour imitates that exhibited by a given natural system [Phillipson & Schuster, 2009]. It means the identification of relevant magnitudes and interactive causal influences among them, as well as the proper description of such influences on the time rate of change of the evolving magnitudes. In fact, the approach is a particular application of the rather general views underlying the fundamental equations of the scientific theories, always covering dynamical contents. The successes achieved have convinced researchers that everything admits to be modelled and initiatives concerning so complex things like a living cell [Karr et al., 2012] or the human brain [Eliasmith et al., 2014] are now under way. At this respect, however, it is worth recalling the well-known case of the Navier-Stokes equation in relation to the mechanical behaviour of fluid flows, for which the success in the behavioural description is in contrast with the lack of explanation of the transition mechanism to turbulence [Frisch, 1995]. It illustrates how, dealing with complexity, the achievement of a model properly imitating the behaviour of a given system does not guarantee an understanding of such behaviour and brings us to consider dubious that, without comprehension of the dynamical mechanisms over which its activity is sustained, any hypothetically successful model for cells or brains could provide answer to so ambitious questions like what is life or how the brain works, although, of course, we don't question the relevance of such models for applied purposes.

The behaviours of complexity arise in systems where a large number of dynamical activities are evolving under enough interactive binding to sustain the system individuality. The tentative explanation of the whole system behaviour would necessarily require some scalable mechanism allowing for the coordinated participation of unlimited numbers of degrees of freedom into a given dynamical scenario, often involving a disparate range of time scales. Under this view we generically inquire what the dynamical systems can do for such a purpose and conclude that they can do just one thing: to oscillate. In agreement with the intuition of Landau when tentatively tried to explain the transition to turbulence like a combination of oscillations [Landau, 1944], any complex dynamical activity, presumably including those occurring in cells and brains too, should be nothing but some kind of complex oscillation and then the problem transforms in what kinds of complex oscillations the differential equations can sustain. To the best of our knowledge, complex oscillations are only achievable through the combination of more simple oscillations and the generalized Landau scenario is the way to achieve it [Rius et al., 2000a, 2000b; Herrero et al., 2012; Herrero et al., 2016, 2018]. In the optimum development of this scenario, each one of a set of fixed points partaking in the same basin of attraction undergoes successive Hopf bifurcations up to exhaust its stable manifold and the resulting periodic orbits participate in nonlinear mechanisms of mode mixing through which the various oscillation modes combine ones within the others and, in particular, most of them manifest together but intermittently in the necessarily recurrent time evolution of the attractor. The peculiar combination of oscillation modes of each phase space trajectory describes the corresponding sequence of dynamical activities, as characterized by their frequency and by their phase-space orientation defining the participation of each variable of the system. Thus, the scenario illustrates how an autonomous system is able to sustain an organized assemblage of dynamical activities by articulating their interactive conjunction into the whole functioning, and it may be expected to be highly scalable concerning the number of combined oscillation modes and the diversity of involved time scales.

The complex oscillations of the generalized Landau scenario have been experimentally demonstrated with a family of physical devices of successively increasing dynamical dimension (up to 6) [Rosell et al., 1991; Herrero et al., 1996; Rius et al., 2000b] and related mathematical models [Rosell et al. 1995; Farjas et al., 1996; Rius et al., 2000a], in which the scenario develops in its simplest form by involving a saddle-node pair of fixed points only. The scenario possibilities through larger sets of fixed points have been tentatively predicted from generic considerations on the presumably involved mechanisms [Herrero et al., 2012]. In this article we present a family of systems of differential equations of arbitrary dimension admitting to be designed such that they are able to develop the oscillatory scenario over an arbitrarily large number of fixed points, and their behaviour is illustrated with some numerical examples dealing with sets of four fixed points.

It is worth stressing that the considered kind of systems has not been inspired by any tentative modelling but by the aim of achieving designable systems under enough control. Although we have tried to maintain the maximum generality, the achievement of such a control has required some restriction in the couplings among variables possibly implying efficiency losses in the oscillatory mixing mechanisms. The design procedure ends with a problem of solving a polynomial system that implies a technical limit in the achievement of designed systems, essentially affecting the number of imposed Hopf bifurcations. Nevertheless, we think that the approach allowed us to penetrating the space of dynamical systems towards appropriate systems for illustrating how the basic features of the generalized Landau scenario around a saddle-node pair extend to larger sets of fixed points and for indicating in this way its scalability possibilities.

## 2. Designable Systems with an *m*-Directional Nonlinear Vector Field

A very general description of $N$-dimensional systems is

$$\dot{z} = Az + \sum_{j=1}^{m} b_j f_j(z, \mu), \qquad (1)$$

where $z \in \Re^N$ is the vector state, $A$ is a constant $N \times N$ matrix, $b_j$ are constant $N$-vectors, $f_j$ are scalar-valued functions nonlinear in $z$, $\mu$ describes constant parameters involved in the nonlinear functions, and the $m \leq N$ components $b_j f_j$ are linearly independent. The multi-directionality $m$ of the nonlinear part of the vector field generically determines the topological structure of the potential sets of fixed points. In effect, it is generically possible to choose a new basis including the $b_j$ and with the rest of vectors, $a_j$, $j = m+1, .., N$, orthogonal to all the $b_j$. By projecting the equilibrium condition, $\dot{z} = 0$, onto the vectors $a_j$ we obtain

$$a_j A z = 0, \quad j = m+1, m+2, .., N,\tag{2}$$

that define a set of $(N-1)$-dimensional hyperplanes passing through the origin. If $A$ is non-singular, the normal vectors $a_j A$ are linearly independent and the intersection of the $(N-m)$ hyperplanes of Eq. (2) reduces to an $m$-dimensional linear subspace, within which the potential fixed points should stay. The actually existing fixed points are determined by the $m$ projections of the equilibrium condition onto the vectors $b_j$ and, under optimum nonlinearities, they may extend like $m$-dimensional arrays so that a basin of attraction may include up to $3^m$-1 saddle points on the separatrix, in addition to the attracting point [Herrero et al., 2012]. The design of systems with $m = 1$ has been reported in a previous work [Rius et al., 2000a] and the method is now extended to arbitrary $m$ values. For this purpose, the generic system (1) should be particularized in different ways.

Firstly, the generality of system (1) is restricted by assuming $N \geq 2m$ and the $m$-dimensional subspace of the fixed points contained within the span of the $a_j$, so that we can choose a basis like ($b_1, .., b_m, a_{m+1}, .., a_{N-m}, e_{N-m+1}, .., e_N$), where the $e_j$ describe the subspace of the fixed points. In a basis like this, the equations may be written as

$$\dot{x}_j = \sum_{q=1}^{N} c_{j,q} x_q + f_j\left(x_q, \mu\right), \quad j = 1, 2, .., m,$$

$$\dot{x}_j = \sum_{q=1}^{N-m} c_{j,q} x_q, \quad j = m+1, m+2, .., N,\tag{3}$$

where the $x_j$ are the components of the new vector state. Each equation of the first group contains one and only one of the nonlinear functions and the absence of the $m$ variables $x_{q>N-m}$ in the right-hand side of the second group implies that the fixed points have the rest of variables equal to zero, provided the corresponding matrix of $c_{j,q}$ coefficients should be non-singular.

Another restriction is done by reducing the presence of the $m$ variables $x_{q>N-m}$ in the first group of equations to only one of them in each equation as follows

$$\dot{x}_j = \sum_{q=1}^{N-m} c_{j,q} x_q + c_{j,N-m+j} x_{N-m+j} + f_j\left(x_{q\leq N-m}, x_{N-m+j}, \mu\right), \quad j = 1, 2, .., m,\tag{4a}$$

$$\dot{x}_j = \sum_{q=1}^{N-m} c_{j,q} x_q, \quad j = m+1, m+2, .., N,\tag{4b}$$

This restriction makes the equilibrium solution of each one of the variables $x_{q>N-m}$ independent of those of the others. It means that the one-dimensional bifurcations leading to the appearance/disappearance of fixed points within the $m$-dimensional subspace should happen just along the $e_j$ directions in the linear regime. It means also that a bifurcation creating new solutions along a given direction will usually produce a multiplicity of fixed points according to the previously existing solutions for the rest of variables and, in this way, the fixed points would appear by forming regular and full multidimensional arrays extended according to the $e_j$ directions.

Finally, the nonlinear functions are particularized by assuming that they are functions of a single variable that, in its turn, is a linear combination of the system variables as follows

$$f_j\left(x_{q\leq N-m}, x_{N-m+j}, \mu\right) = \mu_j\, g_j\left(\psi_j, \mu\right), \quad j = 1, 2, .., m,\tag{4c}$$

with

$$\psi_j = \sum_{q=1}^{N-m} d_{j,q}\, x_q + d_{j,N-m+j}\, x_{N-m+j}\,, \qquad (4d)$$

and where the scale factors $\mu_j$ are introduced to be used as control parameters. As it will be shown, this kind of function allows us to divide the design problem into two separate problems of clear solution: The determination of the coefficients $c_{j,q}$ and $d_{j,q}$ by imposing the occurrence of Hopf bifurcations at a potential set of fixed points, on the one hand, and the choice of the nonlinear functions $g_j$ in order to have the desired fixed points, on the other. The division is possible because the linear stability analysis of the fixed points can be formalized independently of the actual nonlinear functions and, therefore, of the actually existing fixed points. Here resides the trick of the design procedure.

It will be also shown that the procedure has a limit in the number of different Hopf bifurcations to be imposed on a given set of fixed points and this means a limit in the number of free coefficients the system can contain. As it will be seen, such a limit is surpassed in Eqs. (4), provided $N$ is chosen high enough in relation to $m$, and we make now a drastic reduction by fixing the values of all the $c_{j,q}$ of the linear equations (4b) and by choosing 0 for all except for one in each equation. The reduced system is written as

$$\dot{x}_j = \sum_{q=1}^{N-m} c_{j,q}\, x_q + c_{j,N-m+j}\, x_{N-m+j} + f_j\left(x_{q\le N-m}, x_{N-m+j}, \mu\right), \quad j=1,2,..,m, \qquad (5a)$$

$$\dot{x}_j = x_{j-m}, \quad j=m+1, m+2,.., N, \qquad (5b)$$

with the nonlinear functions $f_j$ defined in Eqs. (4c) and (4d). For $m=1$, the free coefficients in Eqs. (5) and (4d) are just what is needed for imposing the maximum number of Hopf bifurcations, while, for $m>1$, there is a deficit growing with $m$ and then successive pairs of additional terms with the corresponding $c_{j,q}$ should be introduced in Eqs. (5b) if more bifurcations are wanted to be imposed. Another peculiarity of the case $m=1$ is that the existence of a linear transformation from (1) to (5) is rather generic since it only requires that the rank of $(b_1, Ab_1, A^2 b_1, .., A^{N-1} b_1)$ be equal to $N$ [Barnett & Cameron, 1985]. Notice that in this case Eqs. (5) are just in the standard form[1].

The systems of Eqs. (4) or (5) may be reinterpreted as a coupled set of simpler subsystems and one of such a kind of decomposition is presented in Appendix A. This kind of view may be useful for the analysis when comparing different systems and their dynamical behaviours, but we find it inappropriate for a good design of the global system through the way of starting from oscillatory optimized subsystems and then increasing the coupling among them.

## 3. Steady-State Solution. Variables and Fixed Points

Systems (4) and (5) have the same steady-state solution. In both cases the equilibrium condition imposes that

$$\overline{x}_{q\le N-m} = 0, \qquad (6a)$$

$$\overline{\psi}_j = d_{j,N-m+j}\, \overline{x}_{N-m+j} = -\frac{d_{j,N-m+j}}{c_{j,N-m+j}}\, \mu_j\, g_j\left(\overline{\psi}_j, \mu\right) = \beta_j\, g_j\left(\overline{\psi}_j, \mu\right), \quad j=1,2,..,m, \qquad (6b)$$

where the overbars denote equilibrium values and the rescaled control parameters

---

[1] It is in principle possible to transform the general system (1) into the standard form with a single nonlinear function appearing in one of the equations only [Gouesbet & Letelier, 1994], but the transformation would be nonlinear, usually implying singularities, and an explicit expression for the nonlinear function of the differential equation would be often unachievable. In any case we find such a transformation not useful for our purpose since it will not provide a nonlinear function of the form of Eqs. (4c-d).

$$\beta_j = -\frac{d_{j,N-m+j}}{c_{j,N-m+j}}\mu_j, \quad j=1,2,..,m, \tag{7}$$

have been introduced to have the solution $\bar{\psi}_j(\beta_j)$ fully independent of the coefficients $c_{j,q}$ and $d_{j,q}$. In Sect. 4, however, it will be shown that the $d_{j,N-m+j}$ cannot be determined by imposing the Hopf bifurcations and then we will choose $d_{j,N-m+j} = -c_{j,N-m+j}$ so that $\beta_j = \mu_j$.

The steady-state solution depends on $N$ and $m$ in the exclusive sense that $(N-m)$ defines the number of variables with null coordinate and $m$ the number of those that can be nonzero, but the solution of each nonzero variable is exclusively determined by the corresponding nonlinear function $g_j$ and the chosen value of $\beta_j$ and, therefore, it is independent of the other variables. In the numerical simulations reported in this article we have considered systems with the same function for the $m$ nonlinear functions and used two different functions:

$$g1_j(\psi_j) = \frac{1.25 - 1.06\cos\psi_j}{1.68 - \cos\psi_j}, \quad j=1,2,..,m, \tag{8a}$$

$$g2_j(\psi_j) = 1.1 - e^{-\left(\frac{\psi_j-10}{2.5}\right)^2}, \quad j=1,2,..,m, \tag{8b}$$

of which the first is periodic and describes the interferometric Airy function of the family of physical devices through which the oscillatory scenario was discovered [Rius et al., 2000b], while the second is simply an inverted Gaussian. For each case, the nonlinear function and corresponding steady-state solution $\bar{\psi}_j(\beta_j)$ are represented in the columns (a) and (b) of Fig. 1, respectively. The steady-state solution is multivalued and then it cannot be analytically expressed, but its graphical representation is achievable by representing on inverted coordinates the single-valued function $\beta_j(\bar{\psi}_j)$ obtained from Eq. (6b). The solution can also be inferred through a graphical analysis of the condition of Eq. (6b) by drawing the straight line $g_j(\psi_j) = \beta_j^{-1}\psi_j$ over the representation of the function $g_j(\psi_j)$, as shown in Fig. 1(a), and by gradually lowering the line slope from the vertical one in correspondence with a $\beta_j$ sweep starting at the 0 value. The intersections between the line and the function define the steady-state solutions of $\psi_j$ for the given $\beta_j$ value and the number of solutions increase in two (or decrease in two) every time a tangency takes place. Every tangency denotes a non-hyperbolic solution and the occurrence of a single-zero eigenvalue bifurcation, typically of saddle-node type[2], like it is the case in Fig.1(a).

The ratio between the slopes of the nonlinear function and the straight line at their intersection is useful for characterizing the steady-state solution of the given variable and to this aim we introduce the following set of auxiliary parameters

$$p_j = \beta_j \left[\frac{\partial g_j}{\partial \psi_j}\right]_{\bar{\psi}_j} = -\frac{d_{j,N-m+j}}{c_{j,N-m+j}} \left[\frac{\partial f_j}{\partial \psi_j}\right]_{\bar{\psi}_j}, \quad j=1,2,..,m, \tag{9}$$

respectively associated with the ratios of slopes for the $m$ nonzero variables of a steady-state solution. Except for the rescaling factor of $\beta_j$, which in practice will be equal to one, the parameter $p_j$ is nothing but the slope of the full nonlinear function $f_j$ at the given $\bar{\psi}_j$ value and, then, the description $p_j(\beta_j)$ of how the $p_j$ value varies upon the steady-state solution $\bar{\psi}_j(\beta_j)$ is also exclusively dependent on the nonlinear function $g_j$, as it is illustrated in Fig. 1(c). Thus, the representations of Figs. 1(b) and 1(c) apply to any system defined by Eqs. (4) or (5) if the involved nonlinear function is that represented in Fig.1(a), independently of the concrete values for $N$, $m$, $c_{j,q}$ and $d_{j,q}$.

---

[2] The transcritical and pitchfork bifurcations are of codimension-one also but require particular conditions on the nonlinear function [Wiggins, 1990] that do not happen in the cases of Fig. 1.

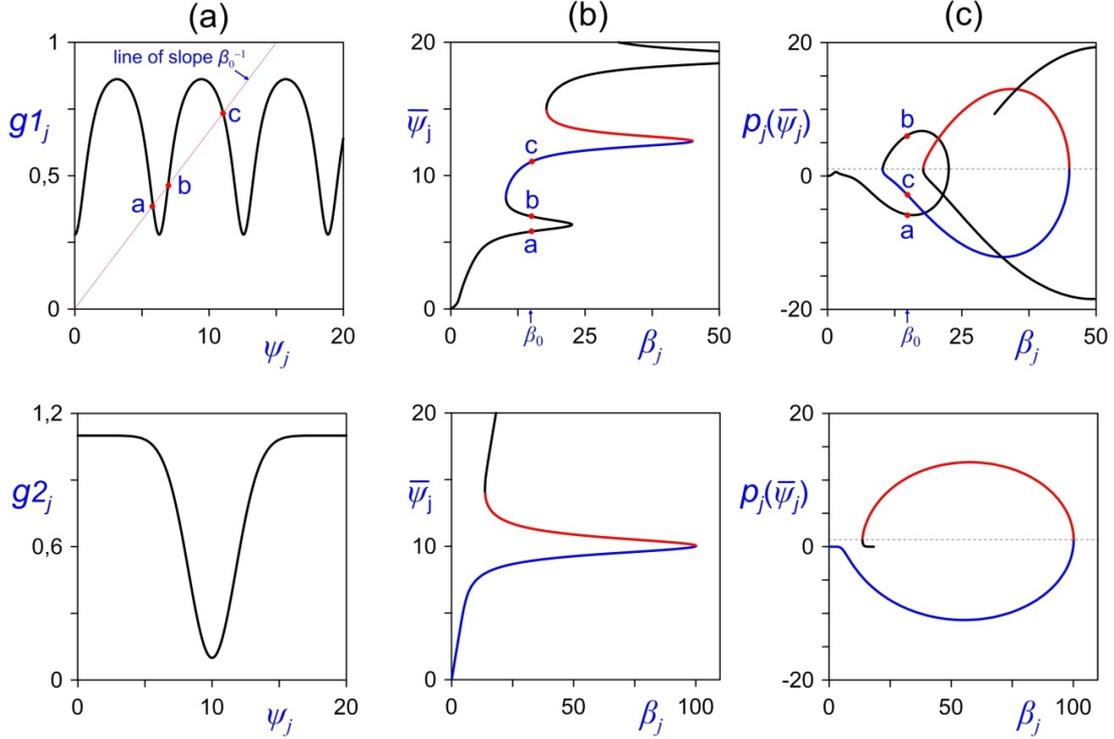

**Fig. 1.** Steady-state solution of one of the $m$ nonzero coordinates of the fixed points, for the two nonlinear functions used in the simulations. (a) Representation of the nonlinear functions $g1$ and $g2$. In the first case the straight line of slope $\beta_j^{-1}$ of Eq. 6b is represented for $\beta_j = \beta_0 = 15$. Its intersections with the nonlinear function denote the steady-state values of $\psi_j$ for the given $\beta_j$. (b) Representation of such values as a function of $\beta_j$. (c) Distribution of $p_j$ values, as defined by Eq. 9, corresponding to the steady-state solution of (b). The dotted line indicates the $p_j = 1$ level at which the saddle-node bifurcations occur. The blue and red curves indicate the steady-state solutions involved in the numerical simulations.

Notice that $p_j$ is equal to 1 for the non-hyperbolic solution of a saddle-node bifurcation and it becomes either higher or lower than 1 for the two solutions emerging after tangency, respectively. Something similar happens in the case of transcritical or pitchfork bifurcations occurring for $g_j$ functions fulfilling the proper conditions. In general, the non-hyperbolic solution always has $p_j = 1$, the solutions existing at both bifurcation sides exchange their $p_j$ value from higher to lower than 1 when crossing the bifurcation, and the neighbouring solutions coexisting at one of the sides alternate their $p_j$ value between higher and lower than 1. The steady-state solutions of a given variable maintain their $p_j$ character of higher or lower than 1 as long as they do not participate in a single-zero eigenvalue bifurcation affecting that variable. Such a character is not altered by the occurrence of one-dimensional bifurcations affecting other variables and, most importantly, it is not altered also by the occurrence of Hopf bifurcations which generically affect all the variables. In the phase space, the multiplicity of coexisting steady-state solutions for the given variable must necessarily alternate their $p_j$ character between lower and higher than 1 (provided $g_j$ is continuous) and such an alternation begins with the $p_j$ <1 of the initial solution, already existing at $\beta_j = 0$ with $p_j = 0$ and whose $p_j$ <1 character remains unchanged while it does not participate in a single-zero eigenvalue bifurcation. On the other hand, from continuity considerations and supposing that no other kinds of limit sets exist, we know that neighbouring solutions should also alternate between attracting and repelling along the $x_{N-m+j}$ direction so that some relation between the one-dimensional stability of the solutions and their $p_j$ character should exist. The $p_j$ value characterizes the feedback effects concerning the given variable only but the stability along the corresponding direction depends also on the rest of variables through the system interrelations. Such feedback influences are common for the multiplicity of solutions and the stability alternation occurs in association with the changes of $p_j$ value from lower to higher than 1. Although the relation of such values with either attraction or repulsion depends on the system of equations, the knowledge of one solution stability would be enough for knowing that of the rest. For instance, if we know that the initial solution with $p_j = 0$ is attractive along the $x_{N-m+j}$ direction, we can then associate the $p_j$ <1 character with attraction along that direction and, vice versa, the $p_j$ >1 character with repulsion. Of course, this stability criterion would be reversed if the initial solution is repulsive along the considered direction.

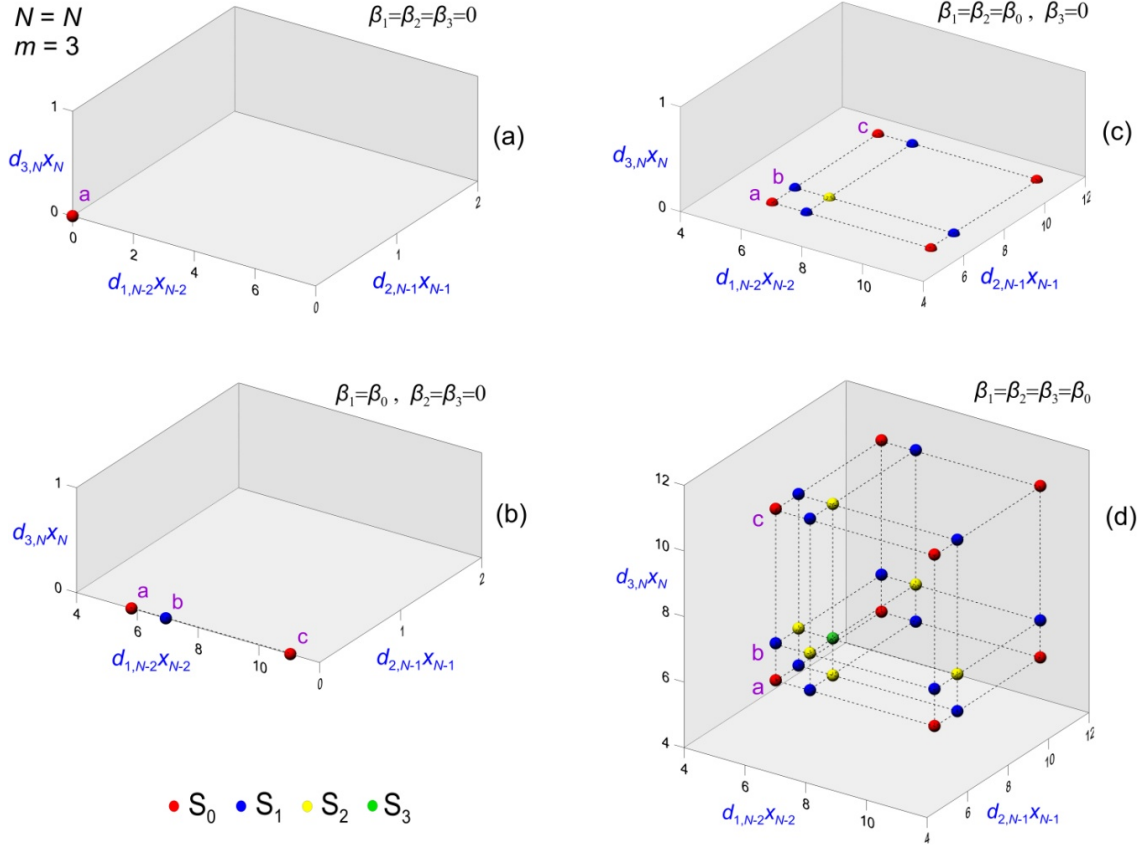

**Fig. 2.** Illustration of how the fixed points appear in the phase space by extending on a regular *m*-dimensional array, for *m* = 3 and arbitrary *N*. The three control parameters are successively swept to produce a saddle-node bifurcation on the corresponding variable, with b and c denoting the new pair of solutions (see Fig. 1). The dotted lines mark the directions along which the saddle-node bifurcations occur and schematically indicate the network of one-dimensional (actually non straight) invariant submanifolds connecting pairs of neighbouring fixed points across the full array, provided that no limit cycles exist. $S_q$ denotes a fixed point with a number *q* of $p_j$ values higher than 1. The initial fixed point for $\beta_j = 0$, $\forall j$, is always $S_0$ and the new fixed points necessarily appear like denoted by the labelling colour. In the absence of limit cycles and supposing the initial fixed point fully stable, the subscripts of $S_q$ indicate the unstable dimensions of the fixed points.

The description above applies to each one of the nonzero variables of the steady-state solution and the next step is to relate the solution of the variables to the fixed points. For instance, Fig. 2 illustrates the situation for *m* = 3 and arbitrary *N*, when the three control parameters are sequentially increased from 0 to the value $\beta_0$ indicated in Fig.1(a) so that each variable experiences a saddle-node bifurcation. The fixed points are represented in the three-dimensional subspace *($e_{N-2}$, $e_{N-1}$, $e_N$)* with rescaled coordinates $d_{j,N-m+j}x_{N-m+j}$ so that they appear located at $(\overline{\psi}_1^i, \overline{\psi}_2^j, \overline{\psi}_3^k)$, with the superscripts denoting the various solutions of each variable. The existing fixed points will correspond to all the combinations of existing solutions for the various variables and, except for the first one, a saddle-node bifurcation affecting a given variable will produce a multiplicity of pairs of fixed points, as many as previously existing points with different coordinates for the rest of variables. The alternation between $p_j$ lower and higher than 1 occur along each one of the $e_j$ directions, starting always from the $p_j$ <1 character of the initial fixed point, and, in the assumed absence of limit cycles, the associative criterion with attraction or repulsion would work for each direction according to the corresponding stable or unstable behaviour of that point for the given direction. In addition, concerning the dimensions outside the *m*-dimensional subspace, all the fixed points will have the same stability as the initial fixed point because the bifurcations yielding appearance/disappearance of fixed points will leave such dimensions unaltered.

To avoid confusion it is worth stressing that the relation between $\psi_j$ and $x_{N+j}$ expressed in the first part of Eq. (6b) applies to the fixed points only since, according to Eq. (4d), $\psi_j$ denotes the combination of system variables constituting the single variable of the function $g_j$. The variables $\psi_j$ have two features making them very convenient as observables: they contain equilibrated contributions of the various oscillation modes and are sensitive to the relative position of the fixed points, and for this reason we usually analyse the system behaviour through them, either the time evolutions or the phase space representations. In such a kind of phase space representation the fixed points appear located just equal as in the representation based on the $d_{j,N-m+j}$ $x_{N-m+j}$ coordinates but the trajectories, including the attractors, could look rather different.

## 3.1 *Identification of fixed points*

We associate each one of the fixed points with the corresponding values of the $m$-tuple $(p_1, p_2, .., p_m)$ and introduce a distinctive classification among them according to if such values are higher or lower than 1, independently of their concrete values. We find also useful to denote by $S_n$ any of the fixed points having a number $n$ of $p_j$ values higher than 1 so that we have $S_n$ points with $n$ varying from 0 to $m$. In particular, the initial fixed point existing at $\beta_j=0$, $\forall j$, is surely of type $S_0$. The identification procedure distinguishes up to $2^m$ different classes of fixed points, among which there are $\binom{m}{n}$ classes of $S_n$ points differentiated by the $p_j >1$ positions within the $m$-tuple, for $n = 0, 1, .., m$.

Let us assume in the following of this subsection that the initial fixed point is fully stable because it is compulsory if we want to guarantee the existence of attractor while varying the $\beta_j$ parameters and because, in fact, it will be one of the aims to be achieved through the system design. Thus, in the assumed absence of limit cycles, we can apply to all the fixed points the associative criterion between $p_j <1$ ($p_j >1$) and attraction (repulsion) along the $x_{N-m+j}$ direction, for j = 1, 2, .., $m$, and, in addition, we also know that they are attractive in the rest of ($N$-$m$) dimensions. The $S_n$ points will then have unstable invariant manifolds of dimension $n$, linearly associated with the set of vectors $e_{N-m+j}$ for which $p_j >1$, and stable manifolds of dimension $N$-$n$ covering the rest of $e_j$ vectors as well as the ($N$-$m$) dimensions outside the subspace of the fixed points. The $\binom{m}{n}$ classes of $S_n$ points will differ in one or more directions of their unstable dimensions.

The system design is based on imposing the occurrence of successive Hopf bifurcations in a set of fixed points including one of each one of the $2^m$ different classes and, for each bifurcation, such an imposition is done by defining the fixed point through the $m$-tuple of $p_j$ values at which the bifurcation should occur[3] and by choosing the oscillation frequency value. The tentative goal is to produce the Hopf bifurcations within the stable manifold of the fixed points, while the initially unstable manifold is wanted to remain unstable since, according to our expectation, it should be relevant for the good working of the oscillatory mixing mechanisms. The optimum scenario would be achieved by exhausting the stable manifold of all the involved fixed points and this defines the limit in the number of different oscillation modes we can impose to our designable systems. Such a limit is given by

$$L_{HB} = \sum_{n=0}^{m} \binom{m}{n} \left\lfloor \frac{N-n}{2} \right\rfloor = 2^{m-1}\left(N - \frac{m+1}{2}\right) \tag{10}$$

where the binomial coefficient describes the number of different classes of $S_n$ points and the integer floor function defines the maximum number of Hopf bifurcations each one of such points can sustain in the ($N$-$n$) dimensions we are assuming for its stable manifold. It is worth remarking that a basin of attraction may contain a higher number of fixed points than the $2^m$ different classes used in the design procedure. In fact, a fully filled basin can contain up to $2^n$ points of each $S_n$ class and this means a total number of $3^m$ fixed points. The coexisting points of the same $S_n$ class will in general have different $p_j$ values so that they will experience the imposed Hopf bifurcations at different values of the control parameters $\beta_j$. Figure 3 illustrates the situation for $m = 2$ and arbitrary $N$ when the parameters $\beta_1$ and $\beta_2$ have been both increased up to produce two successive saddle-node bifurcations in each one of the nonzero variables. The phase space contains 25 fixed points among which there are 4 different classes only: one $S_0$ class, two $S_1$ classes, and one $S_2$ class. According to the assumed stability for the initial fixed point and in the absence of any limit cycle, each $S_0$ is an attractor and the $S_1$ and $S_2$ are saddles of unstable dimension 1 and 2, respectively, located on the separatrix of the various basins of attraction. The central $S_0$ point has a fully filled basin of attraction involving a total number of 9 fixed points, while the rest of $S_0$ points extend unbounded their attraction basin along the plane of the fixed points, as well as all the attraction basins extend unbounded along the additional ($N$-$m$) dimensions. In the $m = 3$ case of Fig. 2(d), the phase space contains the full set of $2^m$ different classes of fixed points: one $S_0$ class, three $S_1$ classes, three $S_2$ classes and one $S_3$ class, and each $S_0$ is surrounded by one fixed point of the other seven classes, but there is no a fully filled basin of attraction, which would include 27 fixed points. The situation of Fig. 2 describes the maximum number of fixed points achievable with the nonlinear function $g2_j(\psi_j)$, Eq. (8b), since it can sustain one saddle-node bifurcation only (and a reverse one), while $g1_j(\psi_j)$ can produce an indefinite succession of bifurcations thanks to its periodicity. At this point it is worth advancing the typical circumstance found in our numerical simulations that, even in case of fully filled basins, efficient nonlinear oscillatory mixing usually happens among the fixed points of one of the basin corners only, involving up to $2^m$ points with one of each different class. Contribution of fixed points of more corners would require additional conditions that we haven't looked for.

---

[3] Although such a determination could suggest a bifurcation of codimension-$m$, it is not the case because, in the space of the dynamical systems, the intersection of the Hopf bifurcation surface with the $m$-parameter family of systems defined by Eqs. (4) (or (5)) with the $\beta_j$ as free parameters would generically be ($m$-1)-dimensional (Subsection 4.1).

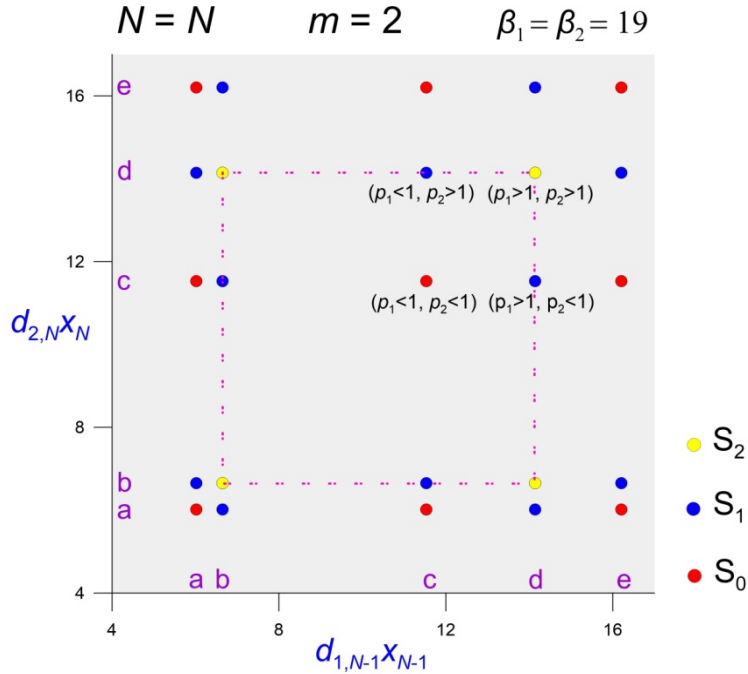

**Fig. 3.** Two-dimensional array of fixed points for $m = 2$ and the nonlinear function $g1_j(\psi_j)$ when two saddle-node bifurcations have occurred on each variable. It contains $2^m$ different classes of fixed points. Assuming the initial fixed point stable and the absence of limit cycles, each of the nine $S_0$ points are stable. The central basin of attraction (schematically marked with dotted line) involves $3^m$ fixed points, the maximum number, although efficient oscillatory mixing usually occurs among fixed points of one of the basin corners only.

## 4. Determination of the $c_{j,q}$ and $d_{j,q}$ through the Linear Stability Analysis

The occurrence of local bifurcations of fixed points, concerning either their own appearance or the creation of periodic orbits, is well characterized by the linear stability analysis. Such an analysis attributes to each fixed point the stability features of the fixed point of the linear system whose matrix of parameters is given by the Jacobian matrix of the original nonlinear system applied to the coordinate values of the given fixed point. The trick of the design method is to be able to use the linear stability analysis without requiring the specific coordinates of any fixed point and to use it to impose the occurrence of successive Hopf bifurcations on each one of the potentially existing fixed points. Of course, the trick is sustained by the chosen kind of system, either (4) or (5), which allows for the association of each one of its potential fixed points with the respective set of $p_j$ values in relation to both its identification and its linear stability behaviour, and this independently of the chosen nonlinear functions because their influences are also contained in the respective $p_j$ values.

The Jacobian matrix of system (4) has the following elements

$$J_{j,i} = \begin{cases} c_{j,i} - c_{j,N-m+j} \dfrac{d_{j,i}}{d_{j,N-m+j}} p_j & , \ i = 1, 2, .., N-m, \\[2mm] c_{j,N-m+j}\left(1 - p_j\right) & , \ i = N - m + j, \\[2mm] 0 & , \ \text{rest of } i \text{ values}, \end{cases} \quad (11a)$$

for $j$ = 1, 2, .., $m$, and

$$J_{j,i} = \begin{cases} c_{j,i} & , \ i = 1, 2, .., N-m, \\[2mm] 0 & , \ i > N-m, \end{cases} \quad (11b)$$

for $j$ = $m$+1, $m$+2, .., $N$, while, in the case of the simplified system (5), the Eqs. (11b) reduce to

$$J_{j,i} = \begin{cases} 1 & , \ i = j - m, \\[2mm] 0 & , \ \text{rest of } i \text{ values}. \end{cases} \quad (12)$$

Thus, the Jacobian elements depend on the system details as implicitly and generically expressed by $J_{j,i}(c_{j,i}, c_{j,N-m+j}, d_{j,i}/d_{j,N-m+j}, p_j)$, where the $p_j$'s describe the fixed point and the $f_j$'s influences on its linear stability. The exclusive presence of the $d_{j,q}$ coefficients through the ratios $d_{j,i}/d_{j,N-m+j}$ results from the definition of the parameters $p_j$ in order to provide them with the well-defined properties they have. Such a kind of dependence means that the linear stability behaviour of the fixed points is independent of one coefficient of each set of $d_{j,q}$ with $j$ = 1, 2, .., $m$, and we fix one of them for each $j$ by choosing $d_{j,N-m+j} = - c_{j,N-m+j}$ in order to have $\beta_j = \mu_j$. This implies a reduction of $m$ free coefficients affecting both system (4) and system (5).

Another restriction of free coefficients is introduced in order to dispose of control on the dissipation degree of the system to be designed. The divergence of the vector field of system (4) is given by

$$\sum_{j=1}^{N} \frac{\partial \dot{x}_j}{\partial x_j} = \sum_{j=1}^{m}\left(c_{j,j} - c_{j,N-m+j}\frac{d_{j,j}}{d_{j,N-m+j}} p_j\right) + \sum_{j=m+1}^{N-m} c_{j,j}, \quad (13)$$

where the $p_j$ denote here a generalization of definition (9) to a generic phase space point and with the last term vanishing in case of system (5). This expression illustrates how the system parameters influence on the degree of dissipation and how it varies through the phase space position according to the set of corresponding $p_j$ values. It applies to the fixed points also and, in particular, to the initial fixed point at $\beta_j = 0$, for which $p_j = 0$, $\forall j$. Thus, the divergence on this fixed point is just the addition of the diagonal coefficients $c_{j,j}$ and, then, this value should be equal to the addition of the $N$ eigenvalues characterizing its linear stability behaviour. Our option is to prefix all the $c_{j,j}$ by choosing a negative value for each one of them and by adjusting it in relation to the highest Hopf frequency. In this way we dispose of control on the dissipation degree of the system and can determine the total amount of the eigenvalues of the initial fixed point, although the possibility of unstable dimensions is not excluded.

The election of the $c_{j,j}$ values implies a reduction of ($N$-$m$) free coefficients, $m$ of which affect both Eqs. (4a) and Eqs. (5a) and the rest affect Eqs. (4b) only. Thus, the number of remaining free coefficients in system (4) is

$$F_{cd}^{(4)} = \left(N^2 - m^2\right) - \left(N - 2m\right), \quad (14a)$$

while for system (5) the number is

$$F_{cd}^{(5)} = 2m\left(N - m\right), \quad (14b)$$

and these numbers should be confronted with $L_{HB}$, the maximum number of Hopf bifurcations the design procedure can impose on the system, as given by Eq. (10). Since at the linear level the imposition of every Hopf bifurcation with a specific frequency will imply the twofold condition of a complex eigenvalue with definite real and imaginary parts, the design problem should be well posed if the number of free coefficients is just twice that of bifurcations to be imposed. For $m = 1$, $L_{HB}$ is equal to $N$-1 and $F_{cd}^{(5)}$ is twice this number so that the free coefficients of system (5) are just what is needed for imposing all the Hopf bifurcations. For $m = 2$, $L_{HB} = 2N$-3, $F_{cd}^{(5)} = 2(2N$-4) and $F_{cd}^{(4)} = N^2$-N, so that system (5) lacks of two coefficients while system (4) surpasses the maximum in a number increasing with $N$. We can design a system (5) by imposing one bifurcation less than the maximum or, alternatively, we can consider an intermediate system between Eqs. (4b) and (5b) by adding two nonzero coefficients and imposing all the bifurcations. For $m = 3$, $L_{HB} = 4N$-8, while systems (5) and (4) can admit $3N$-9 and $\lfloor (N^2 - N - 3)/2 \rfloor$ bifurcations, respectively. In this case, even the full system (4) could lack coefficients if $N$ is not high enough. In general, it may be seen that the condition $F_{cd}^{(4)} \geq 2L_{HB}$ will require $N$ of the order of $2^m$ for counterbalancing the exponential grow in $m$ of the number of disposable fixed points for doing Hopf bifurcations. Nevertheless, for practical reasons, we will be usually unable to impose the maximum number of bifurcations even for system (5) and then the system design will be done by arbitrarily choosing a value for additional pairs of free coefficients.

### 4.1 *Imposition of Hopf bifurcations*

The characteristic equation of the Jacobian matrix is written as

$$\lambda^N + k_1 \lambda^{N-1} + ... + k_{N-1} \lambda + k_N = 0, \tag{15}$$

where $k_q = (-1)^q s_q$, with $s_q$ the sum of all principal minors of order $q$ of the matrix. This equation is universal in the sense that it applies to any fixed point of any continuous dynamical system and both the concrete system and the fixed point under analysis are described through the values of the actual $k_q$ coefficients. We find useful to consider the $N$-dimensional space of $k_q$ coordinates where each point is associated with the corresponding set of $N$ eigenvalues as determined by the characteristic equation. This space contains all the variety of possible fixed points of the $N$-dimensional linear systems and allows visualizing how the design method works to define the system coefficients such that the various fixed points start from the desired $S_n$ types and transform as a function of the control parameters by increasing their unstable dimensions through successive Hopf bifurcations.

The most relevant features of this $k_q$ space are the loci of the nonhyperbolic fixed points, those possessing one or more real eigenvalues equal to zero or/and one or more complex eigenvalue pairs with real part equal to zero, and which may be generically described as $\{0^n, \pm i\omega_1, .., \pm i\omega_q\}$, where $0^n$ denotes $n$ real eigenvalues equal to 0 and the $\pm i\omega_j$ describe a set of $q$ complex conjugate pairs with real part equal to 0 and arbitrary imaginary part. The loci of nonhyperbolic fixed points of a given type defines a geometrical object of dimension $N$-$(n+q)$ that will be denoted like the fixed points and which, for convenience, will be generically termed as a surface. By imposing the corresponding eigenvalue condition on Eq. (15) it is seen that the fixed points $\{0\}$ occupy the $(N$-1)-dimensional subspace $k_N = 0$ and, more in general, the $\{0^n\}$ occupy the $(N$-$n)$-dimensional subspace $k_{N-n+1} = .. = k_{N-1} = k_N = 0$. It is also seen that the fixed points $\{\pm i\omega\}$ form the $(N$-1)-dimensional surface parametrically determined as a function of $\omega$ by either

$$k_1 (i\omega)^{N-2} + k_3 (i\omega)^{N-4} + ... + k_{N-3} (i\omega)^2 + k_{N-1} = 0,$$

$$(i\omega)^N + k_2 (i\omega)^{N-2} + ... + k_{N-2} (i\omega)^2 + k_N = 0, \tag{16a}$$

if $N$ is even, or

$$k_1 (i\omega)^{N-1} + k_3 (i\omega)^{N-3} + ... + k_{N-2} (i\omega)^2 + k_N = 0,$$

$$(i\omega)^{N-1} + k_2 (i\omega)^{N-3} + ... + k_{N-3} (i\omega)^2 + k_{N-1} = 0, \tag{16b}$$

if $N$ is odd. Each value of $\omega$ defines an ($N$-2)-dimensional subsurface that for $\omega = 0$ corresponds to the subspace $\{0^2\}$. With increasing $\omega$, $\{\pm i\omega\}$ emerges from $\{0^2\}$ by developing a complex folding structure with intersections with itself and with $\{0\}$ defining the surfaces $\{\pm i\omega_1, \pm i\omega_2\}$ and $\{0, \pm i\omega\}$, respectively, both of dimension ($N$-2).

As discussed in more detail in Rius et al. [2000a], where illustrative graphical representations are reported, the surfaces $\{0\}$ and $\{\pm i\omega\}$, both of dimension ($N$-1), with their interconnections $\{0^2\}$ and $\{0, \pm i\omega\}$ and the self-intersection $\{\pm i\omega_1, \pm i\omega_2\}$, introduce a partition of the $k_q$ space into $N$+1 different types of regions containing fixed points with a different number of unstable dimensions, from 0 to $N$. The regions with an even number of unstable dimensions appear located in the half space $k_N > 0$, while those with an odd number appear in the half space $k_N < 0$. The contiguous regions at the two sides of $\{0\}$ contain fixed points of unstable dimensions differing in one and the separating hyperplane appears demarcated by $\{0^2\}$ and $\{0, \pm i\omega\}$ in $N$ qualitatively different types of zones where it separates pairs of regions with consecutive numbers of unstable dimensions, i.e., 0|1, 1|2, .., $N$-1|$N$. The surface $\{\pm i\omega\}$ separates regions with unstable dimensions differing in two and it appears demarcated by $\{0, \pm i\omega\}$ and $\{\pm i\omega_1, \pm i\omega_2\}$ in $N$-1 types of zones with different pairs of contiguous regions, i.e., 0|2, 1|3, 2|4, .., $N$-2|$N$.

The first step when analysing the linear stability of a fixed point of a given system is to determine its eigenvalues or, equivalently, to locate it in the $k_q$ space. By modifying the system parameters, the eigenvalues change and the fixed point correspondingly moves through the $k_q$ space, and, if it crosses the surfaces $\{0\}$ or $\{\pm i\omega\}$, a saddle-node or a Hopf bifurcation occurs, respectively, provided the system nonlinearities are appropriate. In a saddle-node bifurcation the fixed point approaches $\{0\}$ while another one approaches also by the opposite side, both become the same nonhyperbolic point and then disappear or, conversely, the contrary process takes place. The involved fixed points possess the same stability except along the direction affected by the bifurcation, and the event happens in the appropriate zone of $\{0\}$ accordingly to the stability of the rest of dimensions. Although less generic, the crossing of $\{0\}$ can also be associated with the transcritical or pitchfork bifurcations by involving different configurations of fixed points in each case. In the crossing of $\{\pm i\omega\}$, the fixed point changes its stability in two dimensions and, under appropriate nonlinearities, a limit cycle emerges around it at one of the surface sides, with a two-dimensional submanifold connecting one another and with the stability of the rest of dimensions remaining unaltered in the fixed point and being inherited by the limit cycle.

We now consider the $m$-parameter families of systems given by Eqs. (4) (or (5)) with a definite set of values for the coefficients $c_{j,q}$ and $d_{j,q}$ but with arbitrary nonlinear functions of the form (4c), and try to visualize where, in the linear regime, their potential fixed points would appear located in the $k_q$ space and how they would move when the $m$ control parameters $\beta_j$ are varied. The $k_q$ coordinates of such fixed points are the characteristic-equation coefficients for the Jacobian matrix (11) (or with (12) instead of (11b)) and are then expressed through polynomial functions implicitly written as

$$k_q = k_q\left(c_{l,i}, c_{j,N-m+j}, d_{j,i}/d_{j,N-m+j}, p_j\right), \quad q = 1, 2, .., N, \tag{17}$$

with the subscripts varying as $l$ = 1, 2, .., $N$ (or $l$ = 1, 2, .., $m$), $i$ = 1, 2, .., $N$-$m$, and $j$ = 1, 2, .., $m$, and where the $p_j$, describing the influences of $\beta_j$ and $g_j(\psi_j)$, are the polynomial variables. The polynomial functions (17) are of degree $m$ but each variable has maximum degree equal to one. In particular, let us explicit the expression for the last coefficient

$$k_N = (-1)^{N(m+1)+m} h(c_{q,i}) \prod_{j=1}^{m} c_{j,N-m+j}(1-p_j), \tag{18}$$

because it allows to verify how any $p_j$ = 1 implies a fixed point $\{0\}$ and, vice versa, any fixed point $\{0\}$ of a system given by Eqs. (4) or (5) should generically have one $p_j$ = 1 at least. In Eq. (18), $h(c_{q,i})$ denotes the determinant of the $c_{q,i}$ coefficients of Eqs. (4b) and it is equal to 1 for system (5). Similarly, from the expressions of $k_{N-1}$, $k_{N-2}$, .., and $k_{N-m+1}$ it is seen that the $m$-tuple ($p_1$, $p_2$, .., $p_m$) with a number $n$ of $p_j$ values equal to 1 should correspond to fixed points $\{0^n\}$.

The $m$-dimensional surface parametrically defined by Eqs. (17), with the $m$ $p_j$'s values as freely varying variables, contains all the potential fixed points of any system with the given set of $c_{j,q}$ and $d_{j,q}$ values, with each one of these points being univocally located accordingly to its values of the $m$-tuple ($p_1$, $p_2$, .., $p_m$). This $m$- dimensional surface extends for the $k_q$ space and generically intersects the $(N-1)$-dimensional surfaces $\{0\}$ and $\{\pm i\omega\}$, where the single-zero eigenvalue and Hopf bifurcations may occur, respectively. Both intersections would generically be of dimension $(m-1)$ pointing out that both kinds of bifurcations remain of codimension-one. In particular, the intersection with $\{\pm i\omega\}$ may be analysed by introducing Eqs. (17) as the $k_q$ coordinates of Eqs. (16) and obtaining a couple of equations with the $m$ $p_j$'s and $\omega$ as variables. These equations characterize the Hopf bifurcation of the potential fixed points of the families of systems with the given set of $c_{j,q}$ and $d_{j,q}$ values but arbitrary nonlinear functions of the form (4c), as it is illustrated in Fig. B1 of Appendix B for a particular case with $m = 2$.

The design procedure consist in determining a set of actual values for the $c_{j,q}$ and $d_{j,q}$ coefficients so that the corresponding $m$-dimensional surface of potential fixed points crosses the $\{0\}$ and $\{\pm i\omega\}$ surfaces in a convenient way. It is tentatively done by imposing a number of punctual intersections between the $m$-dimensional surface and $\{\pm i\omega\}$ as defined by an appropriately chosen set of values for the $m$-tuple $(p_1, p_2, .., p_m)$ and the frequency $\omega$. Imposing such values on the couple of equations describing the intersection, we have a set of couples of polynomial equations with the $c_{j,q}$ and $d_{j,q}$ coefficients as unknowns and, provided the number of equations is equal to that of unknowns, their solution will provide us with families of systems fulfilling the imposed intersection conditions but, as discussed below, we must then verify if they are appropriate for our purpose. As a matter of fact, however, our technical limitations in solving the polynomial systems significantly restrict the attainment of solutions when increasing the number of imposed bifurcations and the degree of the polynomials, in deceiving contrast with the expectable growing of possible solutions (Technical details in Appendix C). In the case of the dynamical system (5), the polynomials are of degree $m$, while for intermediate systems between (5) and (4) the degree is equal to the number of dynamical equations with unknowns, and, of course, the number of different terms in the polynomial equations significantly grows with both $m$ and $N$.

After choosing the values of $m$ and $N$ ($\geq 2m$) and the kind of system between (5) and (4), the design proceeds with the selection of the set of values for the $m$-tuple $(p_1, p_2, .., p_m)$ and the frequency $\omega$ that define the distribution of imposed bifurcations among the various classes of fixed points, and also of the values for the $c_{jj}$ coefficients that characterize the dissipation degree of the system. Initially, the number of imposed bifurcations is taken just the maximum allowed by the number of free coefficients and then we try with a solver of polynomial equations. If no solution is obtained after several trials with different values for the $c_{jj}$ coefficients, the number of imposed bifurcations is reduced in one, usually the fastest, and some arbitrary value is attributed to two arbitrarily chosen unknowns, and the solver is then applied to the reduced system, in which the degree of some polynomial terms has consequently decreased also. If no solution is achieved a new reduction may be done and so on. If there is success, several solutions are often obtained for the same polynomial system and, trying different values on the predefined coefficients, a variety of solutions may be obtained for the same set of imposed bifurcations. Although the procedure involves elements of both handwork and serendipity, the achievement of solutions under the solver limits is relatively easy and in doing this work we have obtained hundreds of them, mostly corresponding to $m = 2$ with $N$ varying from 5 to 8 but also to $m = 3$ with $N = 6$ or to $m = 4$ with $N = 8$[4].

Assuming suitable nonlinear functions, any of the obtained solutions should provide for families of dynamical systems experiencing the imposed Hopf bifurcations and the consequent nonlinear oscillatory mixing, but we want systems fulfilling what we expect to be the most appropriate circumstances for the optimum development of the oscillatory scenario. Our desideratum is that the initial fixed point be fully stable, that the rest of fixed points will appear with $n$ unstable dimensions in accordance with its $S_n$ type, and that, when the various $\beta_j$ will reach their final values, each fixed point has experienced the corresponding Hopf bifurcations within its stable manifold. The initial fixed point stability assures the existence of attractor and is a necessary, though not sufficient, condition for the achievement of the rest of wanted circumstances, and it is easily verified by computing the eigenvalues of the $m$-tuple $(0, 0, . . , 0)$. The solutions are additionally selected

---

by discarding those yielding the *m*-tuple (1, 1, . . . ,1) with some positive eigenvalue and, then, the eigenvalue behaviours of the $2^m$ different classes of fixed points are analysed when their $p_j$ values vary according to a simultaneous sweeping of the several $\beta_j$ parameters so that they pass for the imposed bifurcations[5], with the aim of verifying their occurrence and characterizing how they affect the invariant manifolds, and also with the aim of detecting the fortuitous but frequent occurrence of additional bifurcations, sometimes appropriate for our purposes.

As it may be expected, the distribution of imposed bifurcations among the different classes of fixed points is the critical step for the successful achievement of appropriate solutions, being especially relevant the distribution order of the bifurcation frequencies. We have simplified our search by always choosing $p_j$ values of similar modulus for the imposed *m*-tuples and by usually ordering the occurrence of successive bifurcations on a given class of fixed point according to their frequencies from lower to higher. In practice, we have essentially varied the frequency order among the different classes of fixed points and our trials indicate that certain kinds of distributions easily result in appropriate solutions while others only provide for solutions yielding unstable initial fixed points, as well as there are kinds of distributions that seem unable to sustain any solution. For examples of successful distributions see Appendix C, where the design details of the systems used for the numerical simulations are presented.

## 5. The Nonlinear Functions

By specifying a set of $m$ $g_j(\psi_j)$ functions, we define a particular *m*-parameter family of systems with the $c_{j,q}$ and $d_{j,q}$ coefficients of a given solution and with the $\beta_j$ as freely varying parameters, and whose fixed points will move within the corresponding *m*-dimensional surface accordingly to the excursions of $p_j$ values determined by the nonlinear functions in the $\beta_j$ sweepings. The essential requirement on the nonlinear functions is that they should describe appropriate dips or/and humps to allow for the occurrence of saddle-node pairs of fixed points with proper values of their $p_j$ parameters, while their detailed expression would have a secondary, although of course relevant, influence on the oscillatory behaviour.

As it has been said, the numerical simulations of this work correspond to systems with the same function for the *m* nonlinear functions and we have used two different functions, the *g1* and *g2* given by Eqs. (8a) and (8b), respectively. The periodicity of *g1* provides for a succession of saddle-node bifurcations with the consequent succession of steady-state branches exhibiting successively higher $p_j$ values and then facilitates the achievement of the appropriate $p_j$ values for the oscillatory scenario, although the profusion of fixed points may originate complications. The single dip of *g2* provides for a simple S-shaped solution with the peculiarity of the higher branch having the corresponding $p_j$ value practically equal to zero as a consequence of the function horizontal flatness outside the dip.

Of course, the last step of the design work should be the analysis of the dynamical systems and it suffers also technical troubles, mainly concerning the localization and continuous following of periodic orbits, at least with our numerical tools. The difficulty particularly affects those orbits manifesting strong oscillatory mixing and also some simple orbits acquiring a really large multiplier, and logically it grows with *N* and *m*. The problem grows also with the range of coexisting time scales and it makes convenient some reduction in the wanted diversity of imposed frequencies so useful for the oscillation modes distinction. There is also a certain influence of the chosen nonlinear function and in particular we have experienced less trouble when using the function *g2* and, for this reason, the majority of the reported simulations correspond to *g2*. In the reported phase-space portraits, the continuation difficulties have occasionally impeded us to reach the proper parameter values for certain periodic orbits but, when we do not see a reason for the orbit disappearance, such orbits have been included at lower values of the control parameters to give a more complete view of the portraits. All the reported simulations correspond to the reduced system (5) and with a maximum multiplicity *m* equal to 2, i.e., with sets of four different classes of fixed points.

The final wish in our desideratum is the achievement of appropriate variable interrelations for optimum oscillatory mixing but it is uncontrollable under the design procedure and remains a matter of chance. The properly designed systems should exhibit oscillatory mixing to some extent but a good case requires a number of them and at the end we cannot avoid the suspicion that a better case is even awaiting us. Although we have only analysed the dynamics of a fraction of the designed systems, the most impressive fact is the large variety of pictures with which the oscillatory mixing manifest in the phase space portraits of the $m = 2$ systems and from our experience we can expect that novel pictures will continue to appear by analysing more systems. In the following we begin by briefly introducing the oscillatory mixing phenomena in the simplest circumstance of a saddle-node pair of fixed points for $m = 1$ and remarking some peculiar features that will facilitate the comprehension of the graphical representations for $m = 2$. The different families of systems are identified with a label like, for instance, $N5m1$ X $g1$ directly specifying the values for $N$ and $m$ and the nonlinear function, and with X= A, B, etc., defining the set of $c_{j,q}$ and $d_{j,q}$ coefficients through the tables of Appendix C. None of these systems is analysed in detail but we try to provide a global view of the nonlinear oscillatory mixing possibilities for $m = 2$ although limiting the presentation to the mixing effects on the periodic orbits only. It will be useful to introduce the notation $LC_j^q$ to denote a limit cycle with unstable and stable invariant manifolds of dimension $j$ and $q$, respectively.

---

[5] For instance, in the simplest case of $m = 2$, this would mean sweepings roughly along the four bisectrices of the $p_1$-$p_2$ plane, near which the imposed bifurcations have been always chosen.

## 6. Brief Overview of Oscillatory Mixing in $m = 1$ Systems

For $m = 1$ the $m$-dimensional surface of the fixed points in the $k_j$ space becomes an straight line and its intersection with $\{\pm i\omega\}$ consists in a discrete sequence of points with a maximum number equal to $N$-1, just the same number of imposable bifurcations in a saddle-node pair of fixed points and just the half of the number of free coefficients in system (5), as given by Eqs. (10) and (14b), respectively. Thus, in this case the design is achieved by imposing $N$-1 Hopf bifurcations to the system of Eqs. (5) and the polynomial problem to be solved consists of $2(N$-1) equations of degree one with the same number of unknowns so that its single solution is generically achievable for arbitrary $N$ values. In addition, there is a clear criterion for a successful distribution of the imposed bifurcations between the two classes of fixed points: When ordered according to their frequency from lower to higher, the various bifurcations should be alternatively imposed on $S_0$ and $S_1$ in order to assure the stability of the initial fixed point and the proper occurrence of the successive bifurcations within the stable manifolds of both fixed points. Instead the frequency order in the successive bifurcations of a given fixed point is not so critical. In general, the properly designed systems have[6] $c_{1,q} < 0$, $\forall q$, and the $q$-succession $d_{1,q}$ with alternatively opposite signs ending with $d_{1,N} = -c_{1,N} > 0$. Finally, the standard form of system (5) for $m = 1$ implies two related features both really relevant for the dynamical analysis. On the one hand, since the Jacobian matrix is in companion form, the non-degenerate eigenvalues $\lambda$ fully determine its eigenvectors as given by

$$\left(\lambda^{N-1}, \lambda^{N-2}, ..., \lambda, 1\right) \tag{18}$$

and, in particular, the two-dimensional eigenspace of a Hopf bifurcation with $\lambda_{\pm} = \pm i\omega$ is defined by

$$\left(...,-\omega^6,0,\omega^4,0,-\omega^2,0,1\right),\ \left(...,\omega^5,0,-\omega^3,0,\omega,0\right), \tag{19}$$

so that the limit cycles of different frequencies will appear accordingly oriented and consequently distinguished, independently from which fixed point they arise. On the other hand, the sequential and inverted differentiation relation between successive variables implies that the relative presence of the various oscillation modes in the time evolutions $x_j(t)$, $j = 1, .., N$, increases in proportion to their frequencies each time the subscript $j$ is decreased in one. Thus, $x_1$ optimizes the observation of faster frequencies while they will be practically imperceptible in $x_N$. This fact makes the axes choice strongly influencing in what modes appear more pronounced in the phase space projections.

---

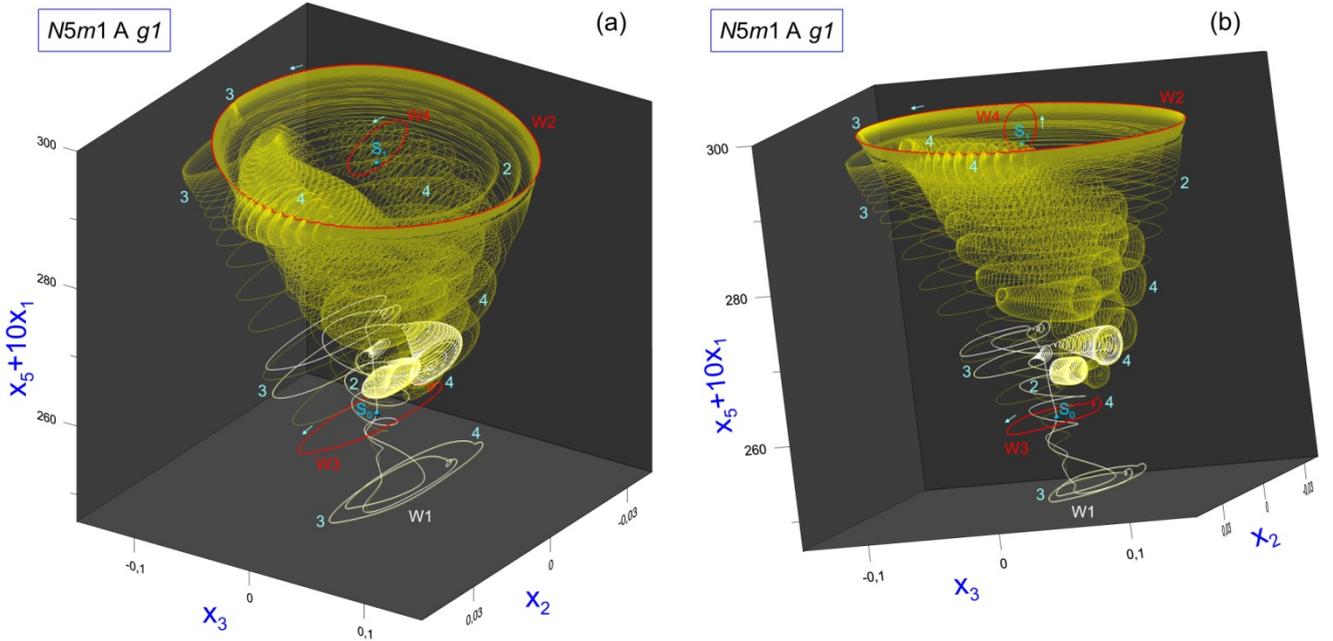

**Fig. 4**. Nonlinear mixing of four oscillation modes emerged from a saddle-node pair of fixed points having both experienced two successive Hopf bifurcations, as exhibited by five particular phase space trajectories: The four periodic orbits, denoted by Wj, and a single trajectory (in yellow) belonging to the two-dimensional unstable manifold of W2 and descending towards the attractor W1. The representation corresponds to $\beta_1 = 24$ and (b) is the same as (a) but with some axes inclination. The clear distinction in orientation of the four oscillation modes (labelled by numbers) makes evident their influence on the represented trajectories: W1 has incorporated localized contributions of the other three modes in several places, W3 has contribution of mode 4, W2 begins to show influences of modes 3 and 4 that will enhance with increasing $\beta_1$, and the yellow trajectory in its way towards the attractor exhibits an extraordinary combination of modes 2, 3 and 4. The time evolutions of the trajectories confirm that each mode contribution has the corresponding frequency.

We are not going to repeat the contents of previous works with $m = 1$ systems and the reader is referred to Rius et al. [2000a, 2000b] and Herrero et al. [2012], where the oscillatory behaviour is mainly investigated through time evolution signals of the attractor, and to Herrero et al. [2018], where a broader view of the scenario is given by showing how the oscillatory mixing may affect extended phase space regions and other periodic orbits besides the attractor. It is worth, however, introducing here a brief overview of the $m = 1$ scenario and we do it with the five dimensional example illustrated in Figs. 4 and 5. The first point to be remarked is that the phase portraits are three-dimensional projections of a higher-dimension phase space so that the trajectories may appear much more intertwined than they actually are and the orientation distinction among the oscillation modes of different frequencies may sometimes appear reduced. Notice also the combination of variables used in the vertical axis of Figs. 4 and 5: $x_5$ is the unique nonzero coordinate of the fixed points, allowing then for their distinction, but it is practically insensitive to the fast oscillations and its combination with $x_1$ provides a better visualization of such oscillations. In fact, we often use the variables $\psi_j$ in the phase portrait representations because each one of them contains contribution of one of the nonzero fixed point coordinates together with an equilibrated contribution of the rest of variables. Look at Table C1 to verify how in the design of $N5m1$A the four bifurcations are alternatively imposed on the $S_0$ and $S_1$ fixed points when ordered according to their frequencies from lower to higher. In this case, and generically for $m = 1$ and arbitrary $N$, the designed systems with such a kind of bifurcation order always create the $S_0$ - $S_1$ pair of fixed points fully stable, the former, and with one unstable dimension, the latter, and, most importantly, the successive bifurcations happen within their stable manifolds so that they increase in two the unstable dimension at each bifurcation. The bifurcations are usually supercritical and, if subcritical, they occur preceded by a cyclic saddle-node bifurcation, the node partner of which is a limit cycle like that that would be directly created in a supercritical bifurcation. Thus, in the case of Fig. 4, the two orbits emerged from $S_0$, denoted by W1 and W3, and those from $S_1$, W2 and W4, have appeared as $LC_0{}^5$, $LC_3{}^3$, $LC_2{}^4$ and $LC_4{}^2$, respectively. The three first orbits have even not experienced any bifurcation, while W4 has become $LC_2{}^4$ at $\beta_1 = 23.6$ by doing a torus bifurcation with a secondary frequency almost equal to that of W2, and W2 is just near to do a period doubling bifurcation, which will occur at 24.026. Of course, the periodic orbits emerge associated with harmonic oscillations at the corresponding frequencies but, with increasing the control parameter, certain of them incorporate influences of other oscillation modes of higher frequency. The incorporation happens in a localized place of the orbit, appearing from nothing and extending more or less along the orbit, sometimes the same mode appears in several places and sometimes several modes appear already combined ones within the others. The mixing process is not related to any instability of the periodic orbit and does not require the creation of new invariant sets but, in the meantime, its dynamical activity initially born as a harmonic oscillation in the Hopf bifurcation has become enriched with the intermittent incorporation of other oscillation modes. Finding rather paradigmatic such a process of becoming complex we introduced the name of *nonlinear complexification* of a periodic orbit to describe it [Herrero et al., 2018]. On the other hand, the trajectory of the W2 unstable manifold in Fig. 4 points out the influence of the several oscillation modes in a wide phase space region and, at the same time, illustrates how complex the time evolution of a single trajectory may become through the oscillatory combination. In fact, the oscillatory influence extends relatively far from the periodic orbits, as it is shown by the transient trajectories of Fig. 5. It is worth remarking that each oscillatory mode appears with the same frequency and phase-space orientation everywhere its influence is manifested in the phase space, by denoting the mode association with a well-defined dynamical activity of the system variables. In addition, although not shown in the figures, a rather relevant feature is that the various oscillation modes with strong mixing among them already appear in the phase space trajectories before the occurrence of the Hopf bifurcations generating the corresponding periodic orbits and even before the saddle-node bifurcation producing the fixed points from which the periodic orbits will emerge (see, e.g., Figs. 1, 4 and 9 of Herrero et al. [2018]). Finally, concerning the potential scalability with $N$ of the $m = 1$ oscillatory scenario, it is worth referring to the simulation for $N = 12$ reported in Herrero et al. [2016] showing how the eleven oscillation modes appear with similar amplitude in the time evolution signal and suggesting then as feasible the absence of a limit for $N$.

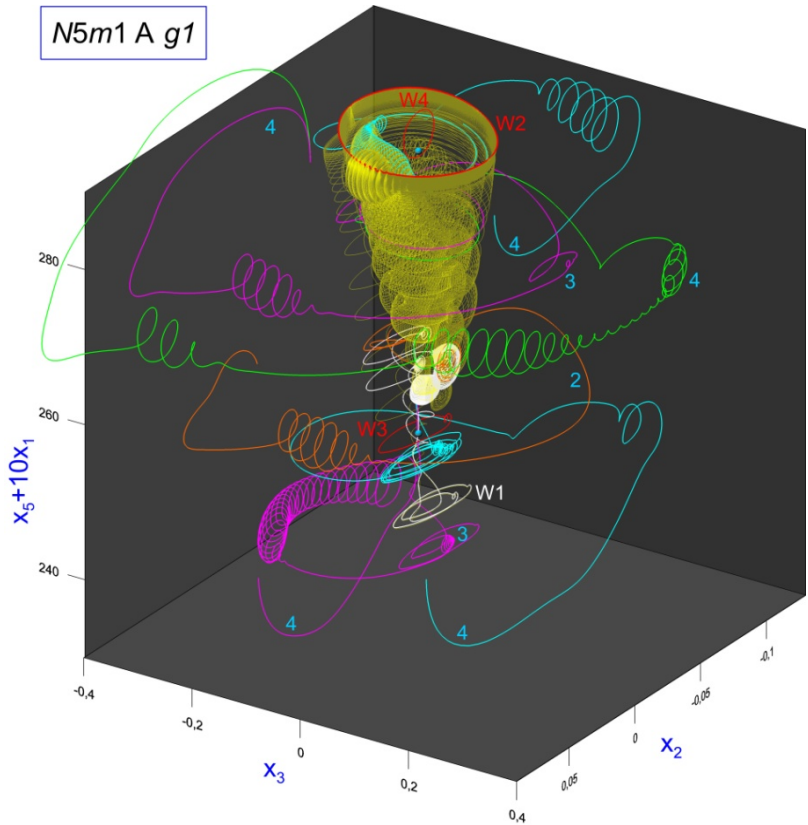

Table 1. Initial points of the transients in Fig. 5

| $x_1$ | $x_2$ | $x_3$ | $x_4$ | $x_5$ |
|---|---|---|---|---|
| 0 | 0 | -0.2 | -1 | 287.92 |
| 0 | 0 | -0.2 | -1.8 | 287.92 |
| 0 | 0 | 0.14 | 1 | 287.92 |
| 0 | 0.06 | -0.15 | -1 | 267.92 |
| 0 | 0.06 | 0.16 | 1 | 247.92 |
| 0 | 0.06 | -0.15 | 1 | 241.92 |

**Fig. 5**. The same as Fig. 4 with a number of transient trajectories initiated (see Table 1) outside the region containing the periodic orbits and the invariant manifolds interconnecting them. The oscillation modes are indicated by numbers and all the helical segments without number correspond to mode 4. All the transients finish on the attractor W1 but their representation has been discretionally truncated. The transient start with a long excursion describing a fast half oscillation is generically found under arbitrary choice of the initial transient point.

We have tentatively tried to explain the oscillatory mode mixing as based on how some periodic orbits extend their oscillations along their unstable manifold towards other periodic orbits like a kind of corkscrew effect [Herrero et al., 2012] (for a descriptive overview see Appendix 3 of Herrero et al. [2016]). This view seems to apply well in a variety of cases but it is not general enough to cover all the circumstances and, of course, it does not apply when the limit sets are lacking. Particularly difficult to be explained in this way is the influence of periodic orbits appeared from the node point towards those emerged from the saddle since the influencing orbits are born with unstable manifolds lacking any connection towards the influenced orbits and its formation during the scenario development seems unlikely. For instance, in Fig. 4 the influence of mode 3 on W2 happens while the three-dimensional unstable manifold of W3 (not drawn in the figure) is fully going towards W1 and, by contrast, the two-dimensional unstable manifold of W2 descends towards W1 by incorporating oscillations of mode 3 (as well as of mode 4) upon its own $\omega_2$ oscillations, apparently suggesting that the influence of mode 3 is going from W2 to W3. On the other hand, the presence of high-frequency modes on the transient trajectories far from the attractor, as it is shown in Fig. 5, cannot be also associated with any unstable manifold of the periodic orbits.

## 7. Nonlinear Complexification of Periodic Orbits in $m = 2$ Systems

The $m = 2$ systems may have four classes of fixed points that we denote like $S_0$, $S_{1a}$, $S_{1b}$ and $S_2$ and whose stable dimensions (assuming properly designed systems) can sustain up to $2N$-3 Hopf bifurcations while the free coefficients of system (5) provide for a maximum of $2N$-4 bifurcations. Nevertheless, as previously noted, we have been usually unable to solve the polynomial system for the maximum of bifurcations and a number of free coefficients have been then arbitrarily valued in the design process (For details see Appendix C). In the $k_j$ space, the intersection of the surface of the potential fixed points with $\{\pm i\omega\}$ is now one-dimensional, as corresponds to a two-parameter family of systems and as may be seen in the example of Fig. B1 for the case $N6m2$ C. This means that a given Hopf bifurcation of a given class of fixed point may occur within one or more continuous ranges of frequencies, sometimes rather broad, accordingly with the corresponding intervals of $p_j$ values. The dynamical equations (5) contain now two separated chains of inverted differentiation relations among the variables $x_j(t)$, along each one of which the relative presence of the different oscillation modes increases in proportion to the respective frequencies each time the subscript $j$ is decreased in two, but the various frequencies may manifest with different strength on the two chains. The Jacobian matrix is no more in the companion form and if, for simplicity, we assume even dimensions only, with $N = 2n$, the eigenvector associated with a non-degenerate eigenvalue $\lambda$ may be written either as

$$\left(h\lambda^{n-1}, \lambda^{n-1}, h\lambda^{n-2}, \lambda^{n-2}, \dots, h\lambda, \lambda, h, 1\right) \text{ or } \left(\lambda^{n-1}, h^{-1}\lambda^{n-1}, \lambda^{n-2}, h^{-1}\lambda^{n-2}, \dots, \lambda, h^{-1}\lambda, 1, h^{-1}\right) \quad (20)$$

where $h$ describes the relation between the two differentiation chains and is given by

$$h(\lambda, J_{ij}) = \frac{J_{12}\lambda^{n-1} + J_{14}\lambda^{n-2} + \dots + J_{1(N-2)}\lambda + J_{1N}}{\lambda^n - \left(J_{11}\lambda^{n-1} + J_{13}\lambda^{n-2} + \dots + J_{1(N-3)}\lambda + J_{1(N-1)}\right)} =$$

$$\frac{\lambda^n - \left(J_{22}\lambda^{n-1} + J_{24}\lambda^{n-2} + \dots + J_{2(N-2)}\lambda + J_{2N}\right)}{J_{21}\lambda^{n-1} + J_{23}\lambda^{n-2} + \dots + J_{2(N-3)}\lambda + J_{2(N-1)}}$$

$$\quad (21)$$

with $J_{1N} = J_{2(N-1)} = 0$ and where the equality of the two expressions arises from the characteristic equation[7]. In the case of a Hopf bifurcation it is seen that, in addition on the frequency, the two-dimensional eigenspace depends also on the set of $c_{jq}$ and $d_{jq}$ coefficients of the system and on the $p_j$ values of the fixed point. Thus, at the linear level, both the frequency and the phase-space orientation of a newborn limit cycle will vary along the bifurcation line of the two-parameter family and, going beyond the bifurcation, the nonlinear effects define its actual features, either those intrinsic to the given oscillation mode or through their modification by mixing influences of other modes. It is then expectable to find significant differences in some periodic orbits between the observed frequencies and those imposed in the system design, and this independently of the possible period enlargement by the intermittent incorporation of other modes. On the other hand, since the nonlinear effects depend on the system variables, the oscillation modes may manifest with slightly different features in the different phase space regions where they actuate. In addition, there is the unwanted restriction for numerical optimization reasons of the diversity range of imposed frequencies and, in total, the consequence is that the mode identification is not always obvious and sometimes becomes difficult and even uncertain. We identify the different oscillation modes with the corresponding periodic orbits that, as indicated in Table C1, are denoted by Wj in the design process according to the corresponding frequencies $\omega_j$ of the imposed Hopf bifurcations ordered from lower to higher. The additional bifurcations occurring without being imposed are denoted by increasing the $j$ index independently of the frequency order. In all the considered systems the lowest frequency is imposed through the $p_j$ values to be the first bifurcation of the $S_0$ class and this means that, in properly designed systems, the relevant attractor will be W1 or some limit set derived from it. On the other hand, the generic mode mixing feature that the influencing mode is always of higher frequency than the influenced one implies that a given periodic orbit may incorporate mixing contributions of higher $j$ modes.

---

[7] The duality of expressions in (20) is relevant for $\lambda = 0$ with either $p_2 = 1$ ($h = 0$) or $p_1 = 1$ ($h = \infty$), respectively.

The reported $m = 2$ simulations correspond to four different designs, one for $N = 5$ and the rest for $N = 6$. The first design is considered twice by alternatively using the two nonlinear functions, $g1$ and $g2$, while the $N = 6$ simulations are done with $g2$ only. Figure 6 presents the phase portraits for $N5m2$ B with the two nonlinear functions at a given value of the $\beta_j$ parameters, both showing the periodic orbits emerged around a set of fixed points including one of each one of the four classes. At the $\beta_j$ values of the simulations, the system with $g1$ possesses 25 fixed points in a configuration similar to that shown in Fig. 3, where the four ones labelled with $c$ and $d$ on both axes are those involved in the phase space representation of Fig. 6(a). Other sets of fixed points may also develop the oscillatory scenario according to their respective $p_j$ values but they are not considered here. The systems with $g2$ possess 9 fixed points in a configuration like that shown in Fig. 2(c), where the four ones labelled with $a$ and $b$ on both axes constitute the basic set for the oscillatory scenario development of Fig. 6(b), as well as for all the considered systems with $g2$. Although the other five fixed points have at least one of their $p_j$ values practically equal to zero, some of them may also experience Hopf bifurcations and develop the oscillatory scenario to some extent. A complete overview of the phase space may be seen in the case of Fig. 11(a), where the phase portrait covers all the fixed points and shows the different oscillatory scenarios developed around them. All the considered systems behave properly in the sense that the fixed points of class $S_j$ appear with $j$ unstable dimensions and experience the successive Hopf bifurcations within their stable manifold by reducing in two its dimension at each bifurcation.

The periodic orbits have been located and their period and set of multipliers determined, with the exception of the attractor W1 whose continuous following initiated near the Hopf bifurcation was interrupted at lower $\beta_j$ probably due to the pronounced mode mixing influences. What is represented as W1 is one cycle of the asymptotic time evolution that, as shown in Fig. 7 for the cases of Fig. 6, looks rather periodic. Nevertheless, the time signal is not strictly periodic since the faster oscillations of successive cycles do not exactly superpose in the phase space representations. Under the diversity of involved timescales, such a loss of strict periodicity may be attributed to the numerical noise arising from the inherent contradiction between the truncation and round-off errors and, in fact, it has been already appreciated in the time signals at lower $\beta_j$ values while the periodic orbit continuation even works and no bifurcation has occurred. On the other hand, the time signal inspection is in this case rather easy and allows us to exclude the occurrence of a period doubling bifurcation up to the represented situation so that chaos seems not feasible. Something similar happens in all the reported phase portraits, i.e., the represented attractor W1 describes one cycle of the asymptotic time signal appearing as practically periodic but with the faster oscillations not exactly equal in successive cycles. In some of them the time signal analysis is not so easy and then the exclusion of chaos is less sure. Nevertheless, it is worth noting that, if chaos would be developed, the chaotic attractor itself and each one of the multitude of coexisting unstable periodic orbits would have an orbit structure of similar complexity like that of the represented W1.

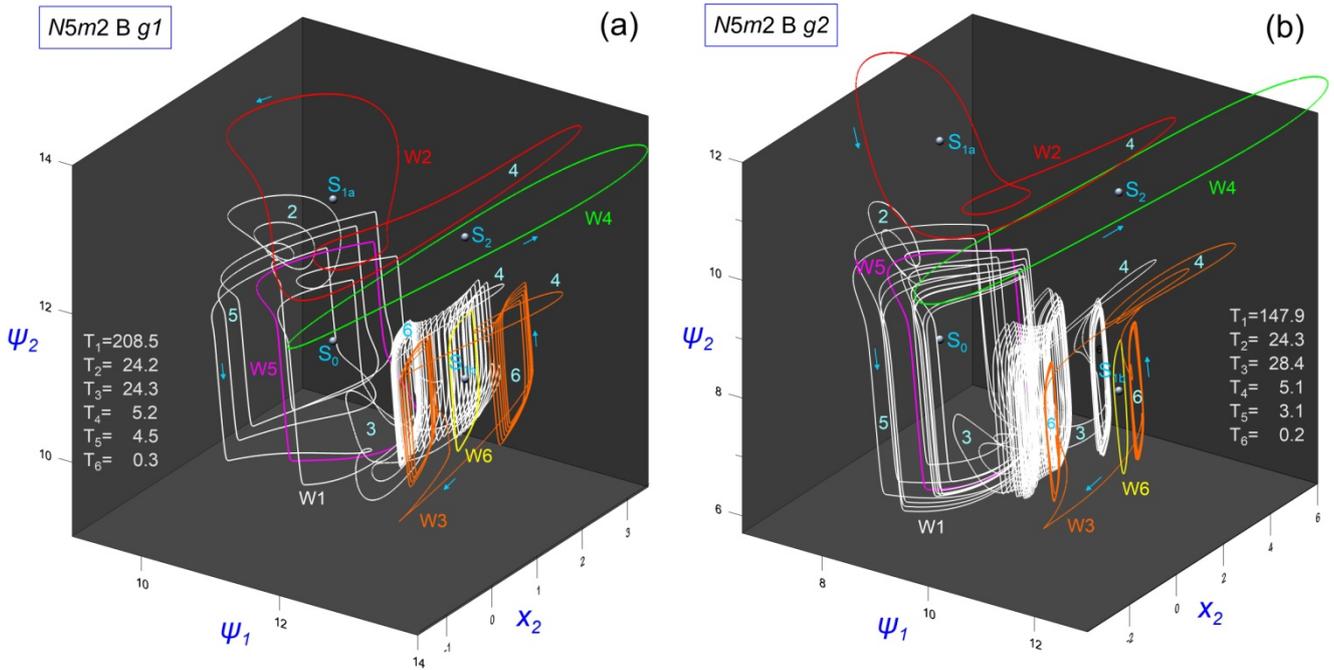

**Fig. 6**. Nonlinear mixing of six oscillation modes emerged from a set of four fixed points in the system *N5m2* B with the nonlinear function *g1* (a) and with *g2* (b), both for $\beta_1 = \beta_2 = 22$, illustrated with the corresponding periodic orbits, denoted by Wj and with their periods indicated. W1 is stable while the rest are saddles. The mixing influences of other modes are denoted by numeric labels.

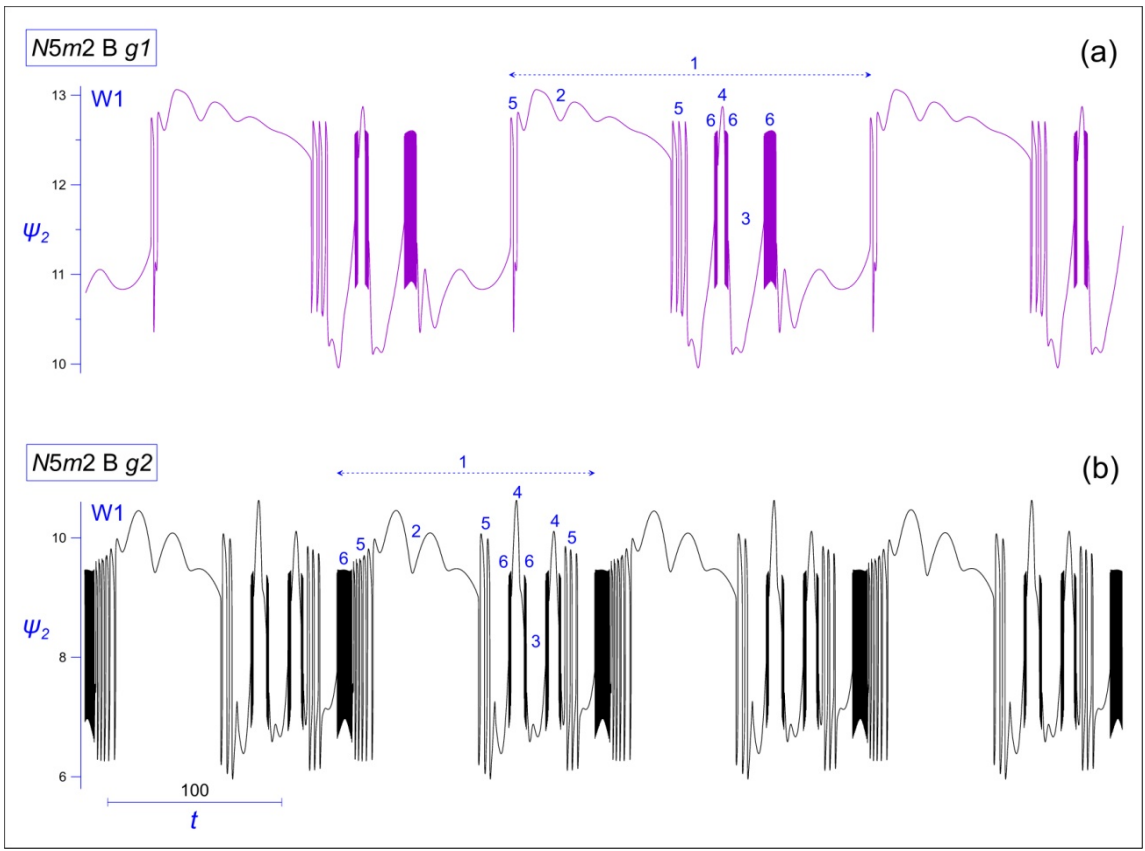

**Fig. 7**.  Some cycles of the asymptotic time evolution on the attractor W1 for the two cases of Fig. 6. The intermittent contributions of the different modes are indicated by numbers.

Notice in Fig. 6 the qualitative similarity of the two phase portraits, with analogous mode mixing influences on each one of the periodic orbits. In fact, the two phase portraits have shown a parallel development of the oscillatory mixing scenario and it is surely related to the fact that the behaviour of the $p_j$ values of the four fixed points as a function of the $\beta_j$ parameters up to the represented situation is relatively similar for the two nonlinear functions, as may be appreciated from inspection of Fig. 1(c). In the two cases, the fixed points have done the maximum number of Hopf bifurcations except $S_{1a}$ that lacks one bifurcation and remains with two stable dimensions. All the periodic orbits have appeared through supercritical bifurcations, W1 like $LC_0^5$, W2 and W3 like $LC_2^4$, W4 and W5 like $LC_3^3$ and W6 like $LC_4^2$, and all of them have developed its oscillatory structure by mode mixing without suffering any bifurcation with the exception of W2 for the system with $g1$ that has experienced a pair of cyclic saddle-node bifurcations after which it has remained $LC_2^4$. The attractor W1 has incorporated intermittent contributions of the other five modes, as it may be appreciated better in the time evolutions of Fig.7. Notice two kinds of mode mixing influences. One at the level of the same fixed point when it has done two successive Hopf bifurcations, like happens to $S_0$ or $S_{1b}$, and the other from top to bottom in the $j$ scale of $S_j$ fixed points, although not within the same $j$ level like between $S_{1a}$ and $S_{1b}$. Notice also the absence of inverted influences from bottom to top in the $j$ scale.

In the phase portrait of the system $N6m2$ C $g2$ shown in Fig. 8 three periodic orbits, W1, W5 and W7 have emerged from $S_0$, two, W2 and W6, from $S_{1a}$, two, W3 and W8, from $S_{1b}$, and one, W4, from $S_2$. All the fixed points have exhausted their stable manifold except $S_2$ which remains with two stable dimensions. The time evolution signals of some of the orbits shown in Fig. 9 aid to interpret their phase-space structure and the W1 portraits for successively increasing values of the $\beta_j$ parameters shown in Fig. 10 illustrate the mode mixing process on this orbit. Notice the presence of an additional oscillation mode that appearing everywhere with a period around 0.004 does not correspond to any of the eight periodic orbits. Such a mode 9 with a frequency around 1570 cannot be associated with the pending bifurcation of $S_2$ since this fixed point remains with two or more stable dimensions in the full two-parameter plane and, in fact, the representation in Fig. B1 of the Hopf bifurcation of all the potential fixed points of the system families $N6m2$ C indicates a maximum frequency of 316 for the bifurcation. Thus the mode 9 bifurcation should be associated with $S_2$ fixed points of neighbouring systems in the space of the dynamical systems, outside of the two-parameter plane.

Besides the ubiquitous presence of mode 9, notice the strong influence of mode 4 on W2 and also, although in minor degree, on W3, being responsible for a significant increase of their periods. W3 receives also influence of mode 8, emerged from the same fixed point, but W8 has incorporated a couple of bursts of mode 9 that become transferred to W3. In its turn, W3 originates the main mixing influences on the attractor W1 by transferring modes 4, 8 and 9 with itself. Figure 10 illustrates how mode 4 actuates in a twofold way on W1 along the $\beta_j$ scale: first on the two lateral sides, in what seems a direct influence from W4 and in which mode 9 also appears, and, second, indirectly through W3. W1 shows a relatively small influence of mode 7 (denoted more by the phase-space orientation than by the frequency) while mode 5 is totally lacking and being these modes emerged from the same fixed point this behaviour is somewhat unusual. Also unusual is the W5 behaviour in the sense that it acquires a rather complex oscillatory structure (Fig. 9 (c)), in which the influence in two places of mode 7 is clear but the other oscillations with periods around 0.015 cannot be identified. Such a complex structure is surely the cause of the W5 continuation troubles and, on the other hand, we can reasonably expect its enhancement to a high degree at the $\beta_j$ values of the phase portrait. W7 shows a small component of mode 9 (not appreciable in the figures). W2 is essentially dominated by mode 4 by extending its period from 20 to the actual 732 but it also incorporates (very short in time) influences of mode 9 alone and of mode 8 with its two mode 9 bursts. If the identification of mode 8 is correct its influence on W2 is rather significant since it would mean mixing between modes emerged from the two classes of $S_1$ fixed points, $S_{1a}$ and $S_{1b}$, and this is far from usual. W6 has a multiplier higher than $4 \cdot 10^5$ making its continuation extremely slow. This periodic orbit contains a small oscillating burst of period 0.012, similar to those unidentified in W5. W4 even maintains a rather harmonic oscillation but at $\beta_1 = \beta_2 = 24$ it begins to incorporate influence of mode 9. Finally, the modes emerged from $S_{1a}$, W2 and W6, have no influence on the rest of periodic orbits.

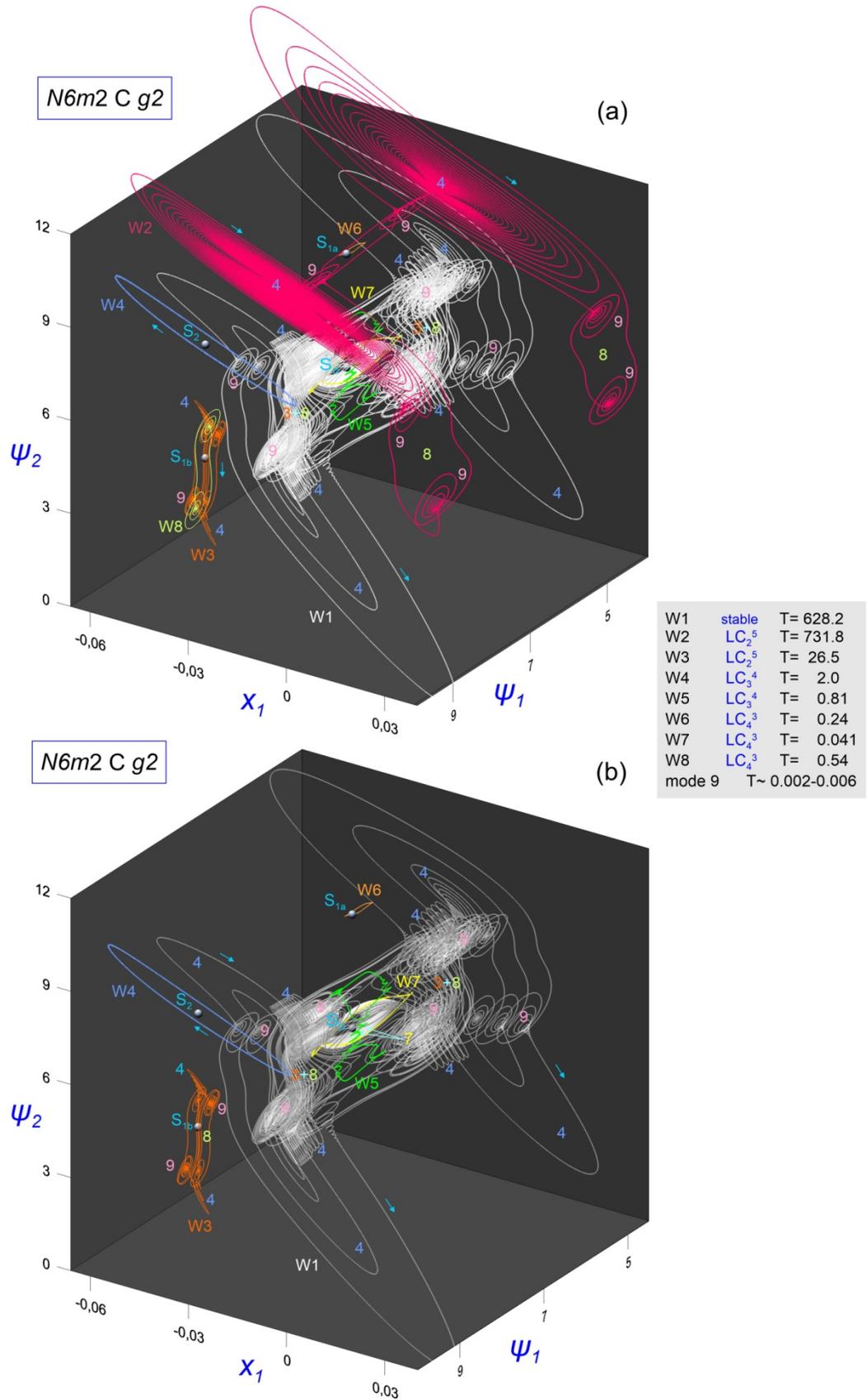

The table within the figure reads:

| W1 | stable | T= 628.2 |
| W2 | $LC_2^5$ | T= 731.8 |
| W3 | $LC_2^5$ | T= 26.5 |
| W4 | $LC_3^4$ | T= 2.0 |
| W5 | $LC_3^4$ | T= 0.81 |
| W6 | $LC_4^3$ | T= 0.24 |
| W7 | $LC_4^3$ | T= 0.041 |
| W8 | $LC_4^3$ | T= 0.54 |
| mode 9 | | T~ 0.002-0.006 |

**Fig. 8.** Nonlinear mixing of nine oscillation modes on the eight periodic orbits Wj emerged from a set of four fixed points in the system $N6m2$ C $g2$, at $\beta_1 = \beta_2 = 20$ except W3 at $\beta_1/\beta_2 = 20/19.1$, W5 at 17.9/18.9 and W6 at 17.6/20. Mode 9 appears everywhere but lacks of periodic orbit. The axes have been adjusted to optimize the visualization ($S_{1b}$ and $S_2$ are at $\psi_1 = 12.1$ and $x_1 = 0$). (b) is the same as (a) but without W2 and W8.

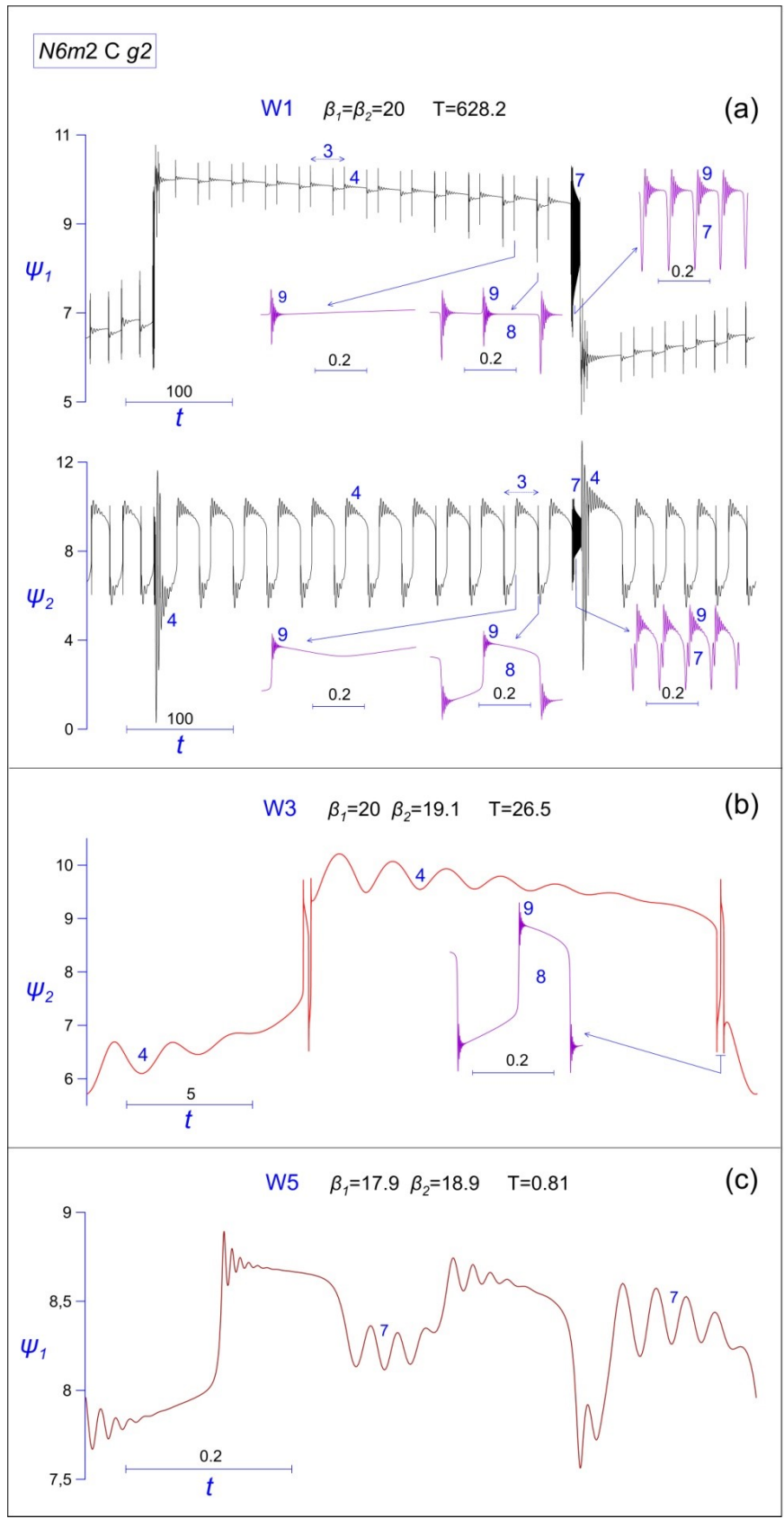

**Fig. 9.** Time evolution signals of one cycle of the attractor W1 and of the periodic orbits W3 and W5 with the contributions of other oscillation modes indicated by numbers. The oscillating bursts without numeric label in W5 are difficult to be associated with any of the other eight modes.

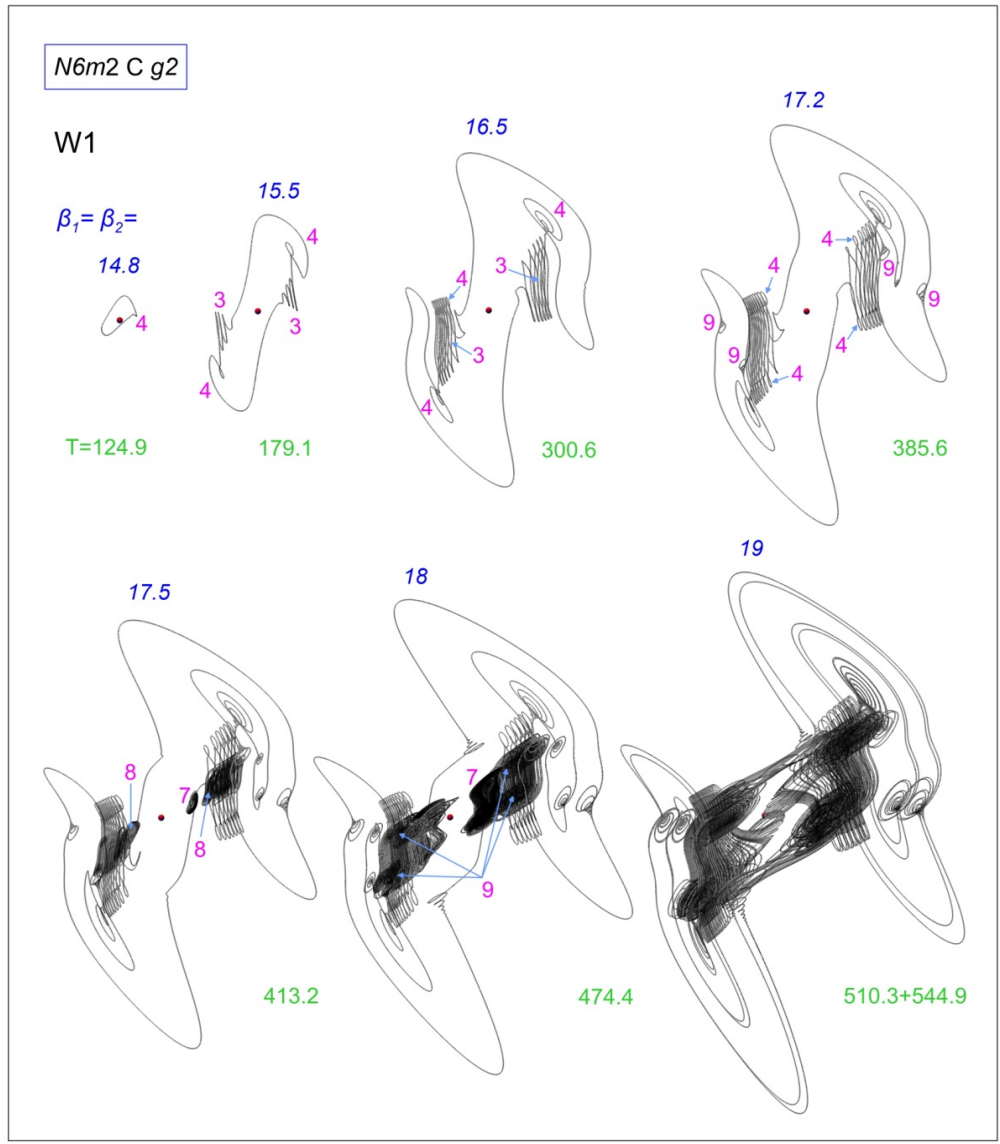

**Fig. 10.** Nonlinear complexification of W1 with increasing the two $\beta_j$ parameters. The successive orbits are represented in the same three-dimensional projection as in Fig. 8 and with the same relative scales on the axes. The first five orbits are located periodic orbits, while the last two describe five cycles of the asymptotic time evolution. The last orbit has a clear period doubled structure. The magenta numbers denote the mixing influence of the corresponding modes as they begin to incorporate on W1.

The bifurcation of periodic orbits is a rare phenomenon in this system, at least up to the represented situation in Figs. 8-10. Along the employed two-parameter continuation paths only W1, W2 and W7 have bifurcated while the rest remained like they were born in the Hopf bifurcation. W2 has suffered two cyclic saddle-node bifurcations after which it has recovered the $LC_2^5$ nature. W7 born like $LC_5^2$ has done two successive torus bifurcations, one the reverse of the other, and a period doubling bifurcation after which it has become $LC_4^3$. W1 born like $LC_0^6$ at $\beta_1/\beta_2 = 14.69/14.69$ has been continuously followed up to 17.5/17.5 without doing any bifurcation and by inspection of the asymptotic time evolution it is seen that a period doubling bifurcation has occurred around 17.6/17.6 but an inverse one quickly takes place so that at 18/18 the signal has become practically periodic again (the represented orbit in Fig. 10 extends over five cycles and although the faster oscillations do not superpose exactly the slow ones do it). Another period doubling happens near 18.15/18.15 and the double cycle structure remains like it may be seen in Fig. 10 at 19/19 but the apparent periodicity is recovered again like it is seen at 20/20 in Fig. 8. Of course, without the orbit continuation we cannot discard the development of chaos but there is no reason to consider such a possibility as underlying the mode mixing process.

Consider now Figs. 11 and 12 dealing with the system family *N6m2* D *g2*. The phase portrait in Fig. 11(a) includes the nine fixed points sustained by *g2* to illustrate how the oscillatory scenario may develop around different sets of fixed points in accordance with how their $p_j$ values vary with the $\beta_j$ parameters. The different fixed points of the same class are distinguished by adding a capital letter into their labels. While each one of the four fixed points in the attraction basin of $S_0$ fulfils well the design conditions of all the imposed Hopf bifurcations, so that they should behave as expected, the other five fixed points have at least one of their $p_j$ values always practically zero and cannot then fulfil the imposed conditions, so that their Hopf bifurcations, if occur, must happen on different parameter values. The periodic orbits W1 and W5 have emerged from $S_0$, W2 and W6 from $S_{1a}$, W3 and W7 from $S_{1b}$, W4 from $S_2$ and W8 will appear also from this fixed point at slightly higher $\beta_j$ values. $S_0$ remains with two stable dimensions and is unable to do its third bifurcation since it requires an enormous $p_1$ value (see Table C1). The initially stable fixed points $S_{0A}$ and $S_{0B}$ have done two and one bifurcations, respectively, and $S_{1aA}$ with initially one unstable dimension has done two bifurcations, while the rest of fixed points do not do any bifurcation. Significantly, the periodic orbits emerged from these fixed points exhibit relatively similar frequencies to those of the Wj orbits and then they are labelled by referring to the Wj of more similar frequency together with the fixed point from which they emerge. It is remarkable how similar some of these periodic orbits look with respect to the relative Wj. Concretely, $W7S_{0B}$ with respect to W7, by including both the mixing influence of mode 8, and $W4S_{1aA}$ with respect to W4, whose influence of mode 8 will appear also on the former at higher $\beta_j$ values, although at the opposite orbit side. Notice however their different stability due to the different one of the fixed points from which they emerge.

The continuation of W1 has been done up to 18/18 and by a coarse inspection of the asymptotic time evolution we consider that it remains practically periodic up to 26/26, after which a period-doubling is apparent. The continuation of W2, W3 and W4 has not reached the $\beta_j$ values of the phase portrait. W2 have not done any bifurcation but one of its multipliers has grown up to 28000. W3 has done a period-doubling bifurcation by becoming $LC_3^4$ and its multipliers remain of moderate value but we were unable to follow it. W4 has done a cyclic saddle-node, a period-doubling and it is approaching another fold bifurcation but one multiplier has grown up to -5390. W5 has done a succession of pairs of fold bifurcations and W6 two torus bifurcations, both having recovered their initial stability status. W7 has not done any bifurcation and W8 will appear in the second Hopf bifurcation of S2 just before 21.5/21.5.

In the one cycle W1 signal shown in Fig. 12(a) it is seen how its oscillatory structure is enriched through mixing influences of modes 5 and 3, both transferring modes 7 and 8 with themselves, and how these two faster modes densely fill a big part of the W1 undulation. Having ordered the successive bifurcations of a fixed point from lower to higher frequency, the mixing influence of the second oscillation on the first one is relatively generic and in the system *N6m2* D *g2* it happens in all the fixed points experiencing two bifurcations. It is also usual that such an influence appears in two opposite places of the first periodic orbit, like happens on W1, W2, W3 and $W1S_{0A}$ but not in W4 and in $W4S_{1aA}$. Significantly, in the $\psi_j$ variables, such a twofold influence appears at intermediate levels of the slower oscillatory undulation, like shown in Fig. 12 for W1 and W3; while, on the other hand, the mixing influences of modes associated with a neighbouring fixed point usually appear on the top or/and the bottom of the $\psi_j$ slower undulation. In Fig.12(b) it is seen that the big bursts of mode 7 on W3 lack of mode 8 but we expect that at the higher $\beta_j$ values of the phase portrait such a mode 8 influence will occur since even W7 lacks of it at the lower $\beta_j$ values of W3. Mode 4 has only influence on W2, while modes 2 and 6 do not participate in any mode mixing except for the influence of the latter on the former. In contrast, the second mode associated with $S_2$, mode 8, has a noticeable influence even before the occurrence of the corresponding Hopf bifurcation, although it always happens in combination with mode 7. Finally, notice the mode 5 influence on W3 that takes place from bottom to top in the *j* scale of $S_j$ fixed points.

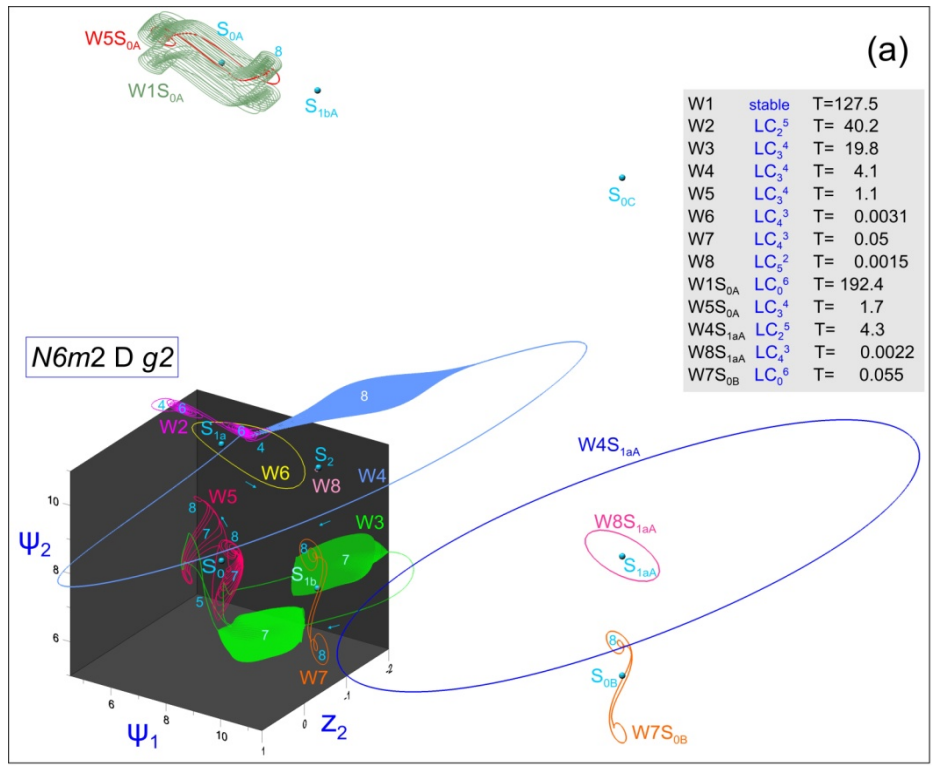

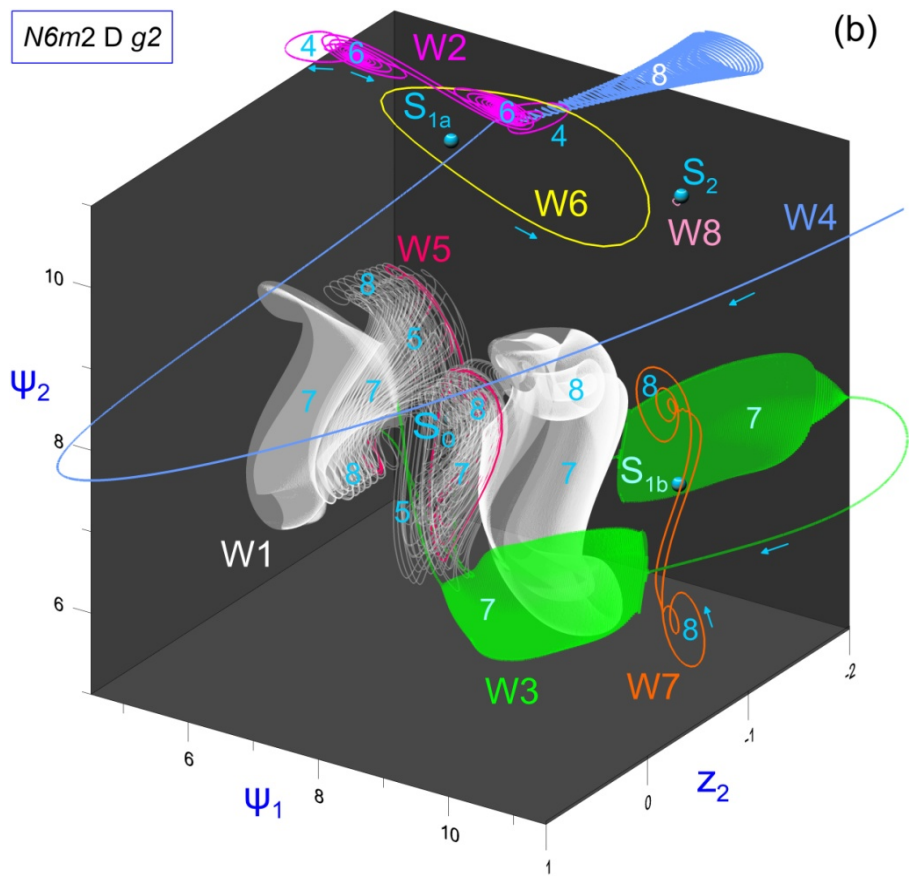

**Fig. 11.** Nonlinear oscillatory mixing in the system *N6m*2 D *g2* at $\beta_1 = \beta_2 = 21$, except W2 at $\beta_1/\beta_2 = 19.5/17$, W3 at 20.1/18.2, W4 at 21/20.7 and W8 at 21.5/21.5. In (a) the phase portrait covers the nine fixed points of the system with the periodic orbits emerged from them, although W1 is not represented for a better visualization of W3 and W5. (b) Enlarged detail around the basic set of four fixed points with W1 included.

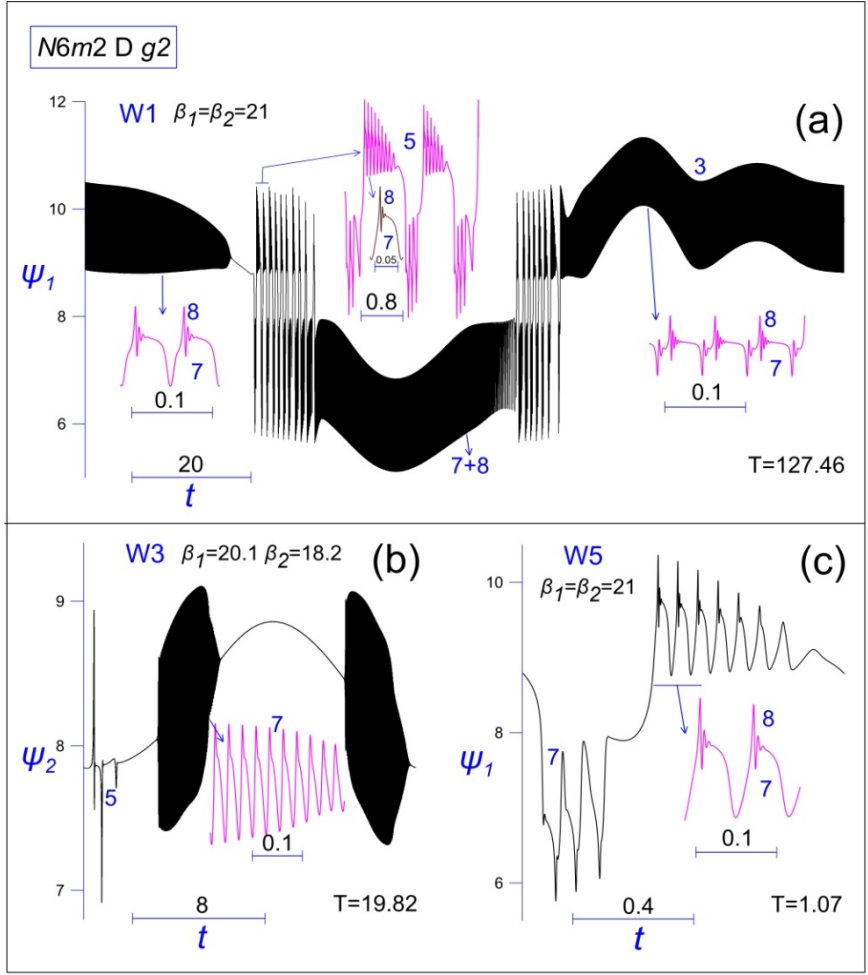

**Fig.12.** Time evolution signals illustrating the oscillatory structure of the practically periodic attractor W1 and of the periodic orbits W3 and W5.

The last selected system is *N6m2* E *g2* and the behaviour of its periodic orbits is illustrated in Figs. 13 to 15. Here again the phase portrait includes more fixed points than the basic set around $S_0$ and illustrates several oscillatory mixing scenarios. We present two different three-dimensional phase-space projections to point out how different may be their visualizations and to remark, in fact, the limits of any projection to provide a faithful description of the six-dimensional phase space. Concerning the Hopf bifurcations, the behaviour of this system is rather similar to that of the previous one. W1 and W5 have emerged from $S_0$, W2 and W6 from $S_{1a}$, W3 and W7 from $S_{1b}$, W4 from $S_2$ and W8 will appear also from this fixed point at higher $\beta_j$ values (24.7/24.7). $S_0$ remains with two stable dimensions and is unable to do its third bifurcation since it requires an enormous $p_1$ value (see Table C1). The initially stable fixed points $S_{0A}$ and $S_{0B}$ have done two and one bifurcations, respectively, and $S_{1aA}$ with initially one unstable dimension has done two bifurcations, while the rest of fixed points do not do any bifurcation.

The continuous following of W1 has been done up to 17.8/17.8 and the asymptotic time signal remains practically periodic up to 25/25. Except W2 and W5, the other periodic orbits have not done any bifurcation along the followed pathways towards the 22/22 of the phase portrait. With W2 we have tried a variety of pathways in the two-parameter plane and verified that it meets in a cyclic saddle-node bifurcation with W5$S_{0A}$, the second periodic orbit emerged from $S_{0A}$. Typically, after emerging from $S_{1a}$ like $LC_2^5$, W2 experiences a significant reduction of its period and an odd number of cyclic saddle-node bifurcations by ending like $LC_3^4$ and, then, it may be moved towards $S_{0A}$ to coalesce on it through a Hopf bifurcation, or vice versa. There are parameter regions with both periodic orbits but only one exists at 22/22 and it is around $S_{0A}$. The behaviour of

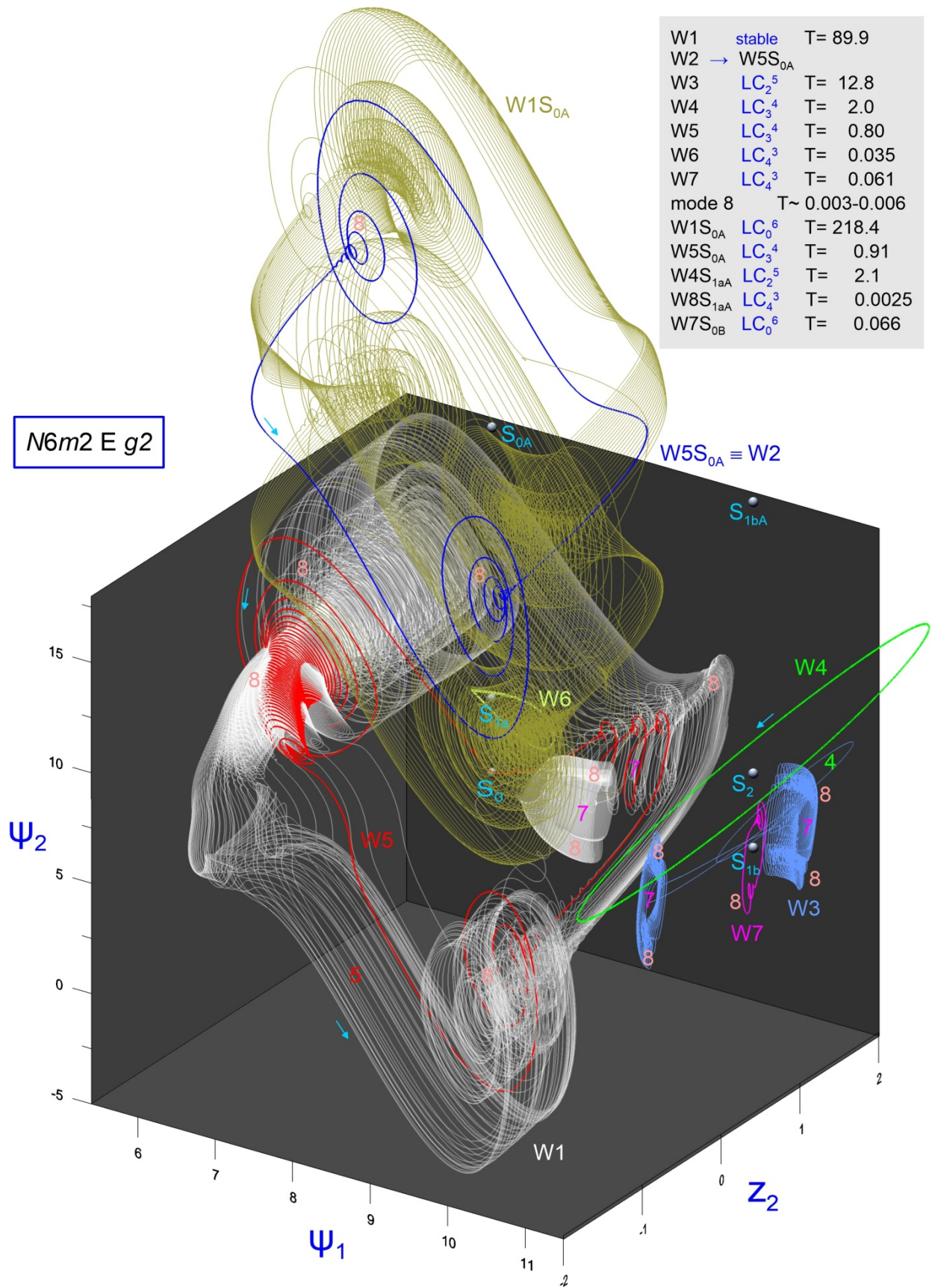

The figure contains the following labels and table:

| W1 | stable | T= 89.9 |
| W2 | → W5S$_{0A}$ | |
| W3 | LC$_2^5$ | T= 12.8 |
| W4 | LC$_3^4$ | T= 2.0 |
| W5 | LC$_3^4$ | T= 0.80 |
| W6 | LC$_4^3$ | T= 0.035 |
| W7 | LC$_4^3$ | T= 0.061 |
| mode 8 | | T~ 0.003-0.006 |
| W1S$_{0A}$ | LC$_0^6$ | T= 218.4 |
| W5S$_{0A}$ | LC$_3^4$ | T= 0.91 |
| W4S$_{1aA}$ | LC$_2^5$ | T= 2.1 |
| W8S$_{1aA}$ | LC$_4^3$ | T= 0.0025 |
| W7S$_{0B}$ | LC$_0^6$ | T= 0.066 |

*N6m2* E *g2*

**Fig. 13.** Nonlinear oscillatory mixing in the periodic orbits of the system *N6m2* E *g2* at $\beta_1 = \beta_2 = 22$, except W6 at $\beta_1/\beta_2 = 19.3/22$. The representation covers six fixed points while the other three, S$_{0B}$, S$_{1aA}$ and S$_{0C}$, are at $\psi_1 = 24.2$. For a better comprehension of the orbit structure of W1 and W1S$_{0A}$ see Fig. 14.

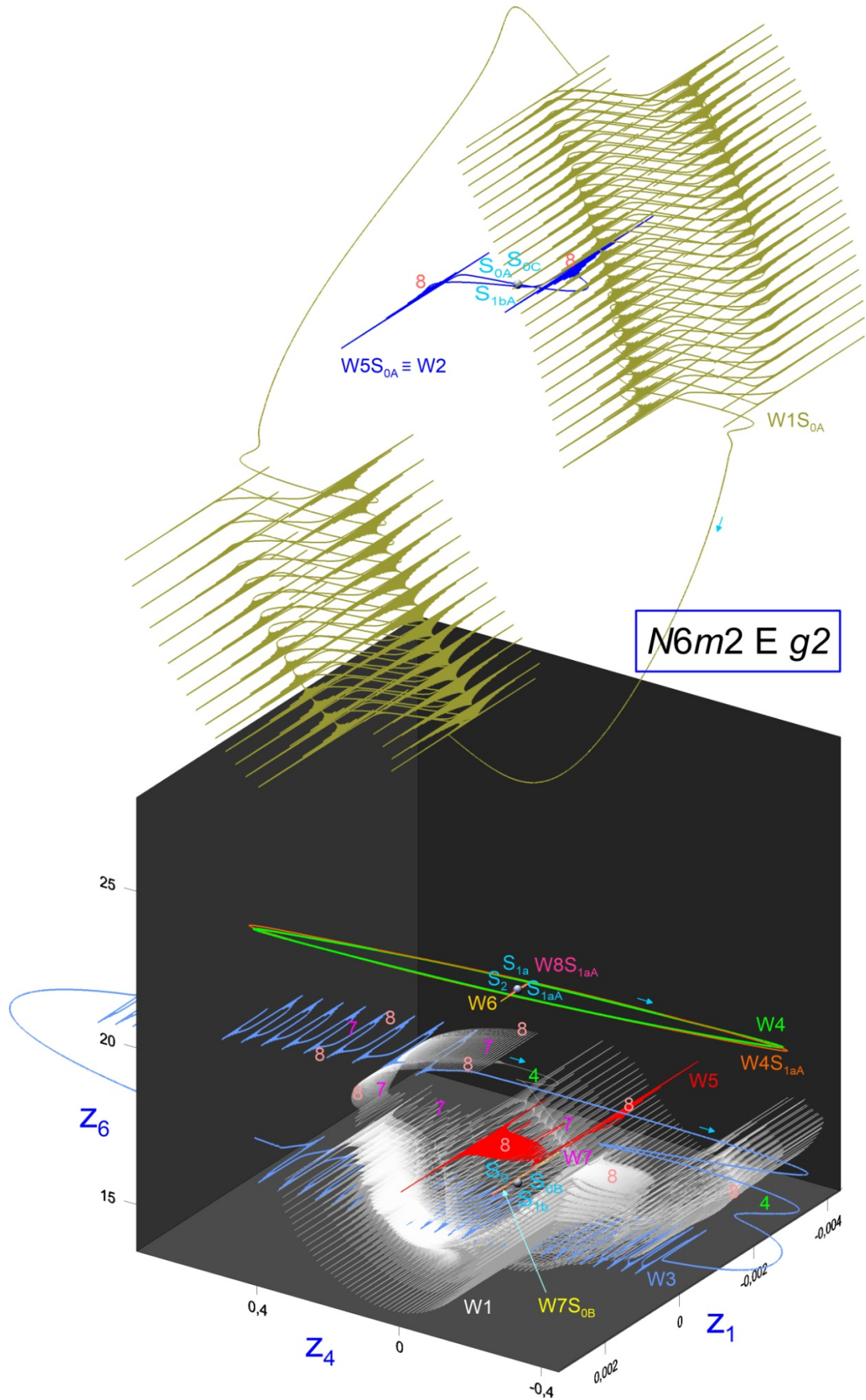

**Fig. 14**. The same as in Fig. 13 but shown in a different three-dimensional projection. The portrait covers now all the fixed points, although distinguished along the vertical axes only, and two additional periodic orbits associated with $S_{1aA}$ and one with $S_{0B}$. W7 and W7$S_{0B}$ appear almost superposed.

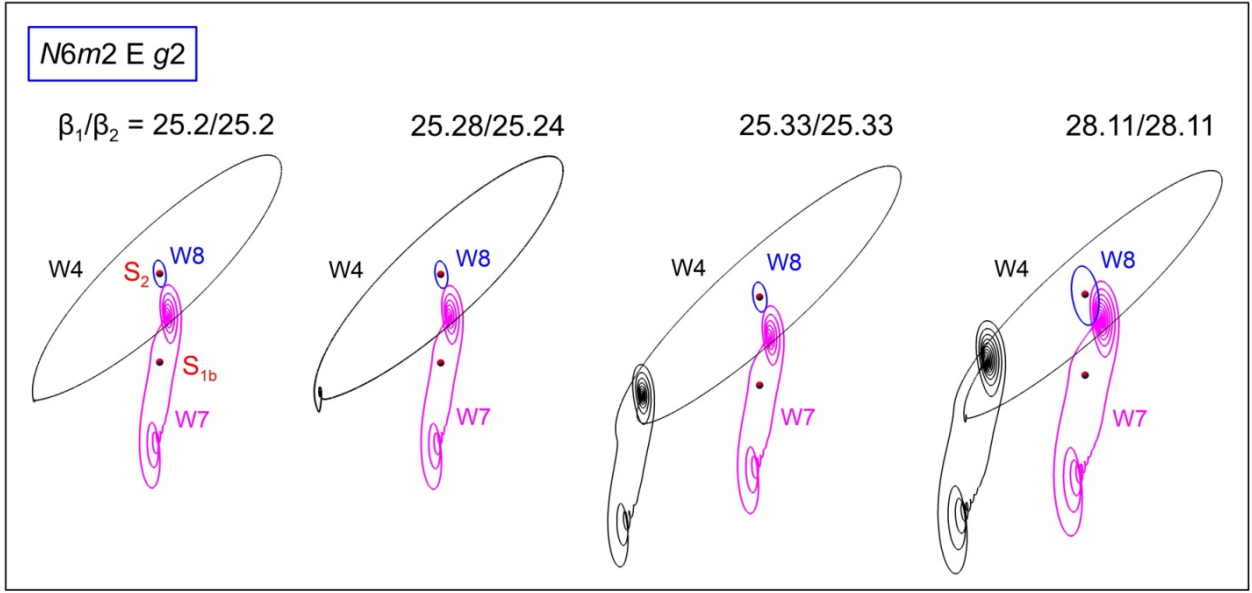

**Fig.15.** Mode mixing process through which the periodic orbit W4 incorporates influence of the orbit W7, from the beginning of the first incorporation to the beginning of the second one. The four representations are on the same scales and the axes are those of Fig. 13.

W5 is also peculiar since after appearing like $LC_3^4$ from $S_0$ with a period of 0.2 at 17.51/17.51, it does not bifurcate and maintains unchanged features while $\beta_2$ is varied up to 22 but with increasing $\beta_1$ it experiences the mixing incorporation of modes 8 and 7 and a huge number of bifurcations: 43 pairs of torus bifurcations, one the reverse of the other and with the orbit becoming $LC_0^6$ in between, combined with 11 pairs of period-doubling bifurcations, one the reverse of the other also and with the orbit becoming $LC_4^3$ in between, and, finally, two cyclic saddle-node bifurcations so that the orbit at 22/22 is again $LC_3^4$. The periodic orbit structure of W1 and of $W1S_{0A}$ is not easily appreciated in Fig. 13 and it may be better understood in the projection of Fig. 14. $W1S_{0A}$ has incorporated in two places a significant mixing influence of the second periodic orbit associated with $S_{0A}$, which in its turn has incorporated influence of mode 8. Similarly, W1 shows the mixing influence of W5 with its incorporated modes 8 and 7 but here the two W5 contributions appear glued together and followed by a long W7 contribution with modes 7 and 8 and one oscillation of mode 4. With increasing the $\beta_j$ values, W1 incorporates additional influences of modes 4 and 3. The period of $W1S_{0A}$ was 193 after the Hopf bifurcation and it has suffered some increase while that of W1 was 100 at birth and has decreased somewhat. The wide presence of mode 8, associated with the second bifurcation of $S_2$ at higher $\beta_j$ values, occurs by influencing W5, W7 and $W5S_{0A}$, all of them the second periodic orbit of the respective fixed points and through them mode 8 reaches into the respective first periodic orbit, W1, W3 and $W1S_{0A}$. There is also influence of W7 with its mode 8 oscillations on W5 and it is transmitted to W1. In addition to modes 7 and 8, W3 incorporates also several mode 4 oscillations but its mode 3 has no influence on other periodic orbits except on W1 at higher $\beta_j$ values. W4 lacks of any influence maintaining an almost harmonic oscillation. Particularly significant is the absence of mode 8 associated with the same fixed point and otherwise of ubiquitous presence in the rest of periodic orbits. Nevertheless, at higher $\beta_j$ values, W4 suffers the peculiar process illustrated in Fig. 15, through which it incorporates influence of W7 with its mode 8 oscillations and which confirms the occurrence of mode mixing from the $S_1$ to the $S_2$ levels. During the represented process and particularly between the second and third portraits, W4 experiences several period-doubling and cyclic saddle-node bifurcations. The S-shaped pathway associated with the latter explains how W4 can significantly transform while both W7 and W8 remain practically unchanged. Like in the case of $N6m2$ D, the two periodic orbits emerged from $S_{1aA}$ are very similar in both frequency and shape to those emerged from $S_2$, being then labelled as $W4S_{1aA}$ and $W8S_{1aA}$, and that emerged from $S_{0B}$ is very similar to the second one of $S_{1a}$, including the twofold mode 8 influence and being then labelled as $W7S_{0B}$. Significantly, at higher $\beta_j$ values, $W4S_{1aA}$ incorporates influence of $W7S_{0B}$ in a similar process to that of W4 shown in Fig. 15, although in this case the mode transmission is from the $S_0$ to the $S_1$ levels. Finally, notice the lack of influence on other periodic orbits of the modes 2 and 6 associated with $S_{1a}$ and the fact that the same happens in the two $N6m2$ cases previously considered but at this respect it is worth mentioning that it does not happen in other designed systems.

Inspection of Tables C1 and C2 shows that the chosen values for the design parameters of the three $N6m2$ systems are very similar. The differences are difficult to be appreciated. First, the seventh bifurcation is imposed on $S_0$ for system C but on $S_{1b}$ for systems D and E. Second, the bifurcations imposed on $S_{1a}$ and $S_{1b}$ have equal frequencies in D and E but different in C. And, third, there are the differences in the values of the $c_{jq}$ predefined coefficients. In contrast, the phase portraits of the three systems look rather different. As a matter of fact, up to date we have (partially) analysed the periodic orbits of 28 $N6m2$ system families covering a wider diversity of designing parameters and it is worth to say that it is difficult to find two of them with similar phase portraits, including cases of multiple solutions for the same designing parameters. Such variability seems due to the multiplicity of possible mode mixing processes and to which of them dominate in each case. In the overall, the reported results point out that the oscillation modes manifest as characteristic ones of the given system by showing relatively well defined frequencies and phase-space orientations, being particularly significant the similitude of periodic orbits emerged from fixed points of different attraction basins.

## 8. Concluding Comment

Let us remark again that we are not trying to model anything but to elucidate the oscillatory possibilities of the dynamical systems in order to sustain dynamical activities of high complexity degree, high enough to admit comparison with those of the activities occurring in living cells, brains or, more simply, turbulent fluids. Essential for this scope is to find a truly scalable dynamical scenario, concerning both the number of degrees of freedom and the range of timescales, and in which an unlimited number of coexisting dynamical activities are able to coordinately evolve. Our work along the years on the generalized Landau scenario has convinced us that it could sustain such a level of scalability and that it could then be a potential candidate for explaining the occurrence of so complex but ordered dynamical activities in the natural world [Herrero et al 2016]. The ultimate aim of this paper is to transmit some of our conviction to the reader and, although it is clear that we have not proved it, we hope that with the overall of reported results on the oscillatory scenario the reader would realize a variety of reasons in favour of its possibilities to be truly scalable. The method of design in itself is defined for arbitrarily large $m$ and $N (\geq 2m)$ allowing, in principle, for the imposition of the maximum number of Hopf bifurcations up to exhaust the stable manifolds of all the $2^m$ classes of fixed points. Certainly, the existence of a solution for the optimum design is not guaranteed even with the full system of dynamical equations (4) since it will depend on the intersection possibilities of the $m$-dimensional surface of the potential fixed points with the surface $\{\pm i\omega\}$ and such possibilities are difficult to be characterized in the general case. It is clear that the highest number of imposable bifurcations yielding a consistent polynomial system will depend on the chosen values for the predefined designing parameters but, leaving our limited ability in finding the solution aside, we take for granted that such a number will increase with increasing $N$ and $m$ so that the number of involved oscillation modes will consequently grow.

Another thing is to attain the good working of the nonlinear mode mixing mechanisms or, in other words, to what degree the nearby occurrence of Hopf bifurcations in the set of fixed points guarantees the appropriate unfolding of the oscillatory mixing scenario. Besides, it is worth recalling here the reduction of interrelations among variables introduced into the dynamical equations to make them designable and the possibility of consequent limitations in their dynamical behaviour. As already said, for $m = 1$ the optimum design with $N$-1 bifurcations is surely achievable for arbitrary large $N$ values with the reduced dynamical system (5) and, according to our experience, the occurrence of the bifurcations in the saddle-node pair of fixed points is enough for assuring the full scenario development in its simplest form, although we have not analysed what happens when the separation of the several bifurcations within the $\beta_j$ parameter plane is increased with the chosen $p_j$ values. Dealing with $m = 2$ systems we have seen how the design procedure opens a wide variety of possibilities when imposing the Hopf bifurcations to sets of four fixed points, but with the drawback of lacking a definite control on how the mode mixing mechanisms work and therefore on how they could be optimized. The reported results illustrate different kinds of mixing processes among oscillation modes associated with the four classes of fixed points and we can affirm that the variety of processes significantly enlarges when considering more system families, but in general the obtained scenarios seem even far from fully developed in the sense of achieving the combination of all the oscillation modes in the attractor. In any case, we consider that the observed variety of processes confirms the scenario extension to some extent and we reasonably expect that $m = 2$ systems developing more complete scenarios must exist. Although designed by imposing noticeably lower numbers of bifurcations than the maximum allowed, we consider feasible the analysis of $m = 3$ systems. Of course, in comparison with the m=2 case, we should expect a more pronounced opening of possibilities and the consequent necessity of a larger number of analysed systems (and a great amount of chance) to catch good enough systems for our purpose. At the end, what we need to verify is the occurrence of mode mixing processes among the 9 classes of $S_j$ fixed points, particularly those involving the higher $j$ classes and sustaining the hierarchical transmission along the $j$ scale towards the attractor. For higher $m$ values the numerical research will become more difficult and in order to clarify the scenario scalability would probably be more useful to advance in the mathematical characterization of the oscillatory mixing mechanisms and particularly of their limiting constraints, by expecting that the understanding for $m = 1$ and $m = 2$ systems will enlighten the extension to the general case.

In principle, since the oscillatory scenario develops in association with the successive Hopf bifurcations of the fixed points, a mathematical theory properly covering it would possibly require to consider the role of the degenerate bifurcations in which all the Hopf bifurcations of a fixed point occur simultaneously and which may be a source of other kinds of bifurcations, and this for the different classes of fixed points, but, at the same time, the fact that the periodic orbits become complex by mode mixing without sustaining instabilities and without requiring the creation of new invariant sets indicates that the secondary bifurcations are not essential for the scenario unfolding. This view is also sustained by the robustness of the scenario unfolding manifested by the continuous and smooth development of the mode mixing processes under variation of any of the involved parameters, either the $\beta_j$, or the $c_{jq}$ and $d_{jq}$, or the ones of the nonlinear functions. The mixing mechanisms should be related to how each oscillation mode expands its influence for the phase space and how the several modes can intertwine their influences to appear together on the trajectories, either transients o limit cycles, and this even before the occurrence of the corresponding Hopf bifurcations since the limit cycle denotes more the culmination than the origin of the oscillatory emergence. Nevertheless, it seems clear that the constraints of intertwinement should be of topological nature and that they will particularly manifest over the unfolding of the several limit sets with their invariant manifolds of varied dimensions sustaining a mesh of connections among them under the no crossing restriction. Significant information could be achieved through the numerical analysis of such invariant manifolds and, more in general, through the analysis of transient trajectories which, although cumbersome, is reasonably feasible for not excessively high dimensions. Without having done any kind of systematic analysis we have noted two peculiar features of the transients in $m = 1$ systems. First, the intriguing fact that, after arbitrary choice of the initial point in regions far from the attractor, the transient generically begins with a fast oscillation (usually the fastest), like may be seen in Fig. 5, where the farther the initial point the longer the path of the first half oscillation. Second, in the approach towards the attractor the transient often describes a succession of oscillatory bursts of different frequencies ordered from higher to lower (see Figs. 4 and 9 of Herrero et al. [2018]).

In the meantime, we find reasonable to expect that the criterion of considering Hopf bifurcations within the stable manifolds only would make compatible the scenario unfolding with the topological constraints and that, at least, a significant part of the allowed maximum number of oscillation modes could appear together on the attractor. Of course, after supposing the oscillatory scenario highly scalable, a tentative explanation of how the natural world might have been able to exploit its possibilities is required in order to make feasible its potential relation with the highly complex things of nature, and it should be through a proper evolutionary mechanism. With this purpose, we have tentatively introduced a framework of structural evolution in dynamical systems based on the optimization of the oscillatory capabilities [Herrero et al., 2012] for which the continuous extension of oscillatory systems within the space of the dynamical systems and the early appearance of faster modes in the phase space would be rather convenient features.

### Acknowledgments

The paper is dedicated to Alexander S. Mikhailov, a former editor of *Physica D* to whom we are deeply indebted because without his fine work as editor our research on the nonlinear oscillatory mixing should have been truncated nine years ago.

## Appendix A

### The $m = m$ System as a Set of $m$ Coupled $m = 1$ Subsystems

To visualize the system of dynamical equations as a set of coupled subsystems we consider the reduced system of Eqs. (5), with the nonlinear function (4c), and assume the simplest circumstance of $N = nm$, with $n$ an integer. The global system may then be seen as a coupled composition of a number $m$ of $m = 1$ subsystems of dimension $n$, with the subsystem $j = 1, 2, .., m$ written as follows

$$\dot{y}_{j,1} = \sum_{q=1}^{n} l_{j,1}^{j,q} \ y_{j,q} + \mu_j g_j(\psi_j) + \sum_{\substack{i=1 \\ i \neq j}}^{m} \sum_{q=1}^{n-1} l_{j,1}^{i,q} y_{i,q} \ ,$$

$$\dot{y}_{j,p} = y_{j,p-1} \ , \qquad p = 2,..,n,$$

with $\ \psi_j = \sum_{q=1}^{n} h_j^{j,q} y_{j,q} + \sum_{\substack{i=1 \\ i \neq j}}^{m} \sum_{q=1}^{n-1} h_j^{i,q} y_{i,q}$

and the relation with the original system given by

$$y_{j,q} \equiv x_{j+(q-1)m} \ , \qquad l_{j,1}^{i,q} \equiv c_{j,i+(q-1)m} \ , \qquad h_j^{i,q} \equiv d_{j,i+(q-1)m} \ ,$$

with the two new sets of coefficients describing the influence of the variable $q$ of subsystem $i$ either on the time rate change of the variable 1 of subsystem $j$ or on the single variable of the nonlinear function of subsystem $j$, respectively.

## Appendix B

## Hopf Bifurcation of the Potential Fixed Points of the System Families defined by a Set of $c_{i,q}$ and $d_{i,q}$ Coefficients

As discussed in Subsect. 4.1, after choosing a set of values for the $c_{j,q}$ and $d_{j,q}$ coefficients, the $m$-dimensional surface defined in the $k_q$ space by Eqs. (17) with the $m$ $p_j$'s as free parameters contains all the potential fixed points of the $m$-parameter families of systems given by Eqs. (4) (or (5)) with the definite set of coefficients but with arbitrary nonlinear functions of the form (4c), and the surface intersection with $\{\pm i\omega\}$ comprises all the potential nonhyperbolic fixed points with a pure imaginary pair of eigenvalues of such systems. It may be expected that the involved nonlinearities generically guarantee the association of the $\pm i\omega$ eigenvalue condition with the occurrence of the Hopf bifurcation and the consequent generation of a limit cycle around the fixed point, and then the intersection with $\{\pm i\omega\}$ describes the potential Hopf bifurcation of the system families with the given set of $c_{j,q}$ and $d_{j,q}$ coefficients. With a specific set of nonlinear functions, the actual $p_j$ excursions with varying the control parameters $\beta_j$ will be accordingly limited and therefore the achievable fixed points also. Concretely, the numerical results reported in this work involve fixed points with $p_j$ values of modulus lower than 10, although the periodic function $g1$, Eq. (8a), may sustain larger values.

The intersection with $\{\pm i\omega\}$ may be analysed by introducing Eqs. (17) as the $k_q$ coordinates of Eqs. (16) to derive a couple of equations with the $m$ $p_j$'s and $\omega$ as variables that generically define a surface of dimension ($m$-1). For $m = 2$ and by taking $\omega$ as a parameter, the intersection equations may be solved and a twofold solution for $p_1(\omega)$ and $p_2(\omega)$ is obtained. For instance, Fig. B1 represents the real solutions for the families of the case $N6m2$ C (Tables C1 and C2). In this case the twofold solution covers finite frequency intervals in the limits of which the two solutions become equal (and complex conjugated at the other side), but there are cases the solutions of which cover all the frequencies in continuity from 0 to probably infinity, as it is the case for $N6m2$ E. Nevertheless, for frequencies very much higher than those imposed in the system design the solutions often correspond to enormous $p_j$ values. At $\omega = 0$ always exist a degenerate twofold solution with $p_1 = p_2 = 1$ that in the case of Fig. B1 remains isolated but in other cases it may also extend along some frequency interval. A vertical line within a dispersive curve is a typical feature of such a kind of representation, appearing several times in all the analysed cases, and it seems really be a vertical straight line reaching infinite values (at least, we have been numerically unable to find their limits). It would mean that the $\pm i\omega$ eigenvalue condition at the given frequency is fulfilled independently of the $p_j$ parameter varying along the line. Its explanation requires realizing that, for arbitrary $m$ and as a consequence of the maximum degree equal to one of the polynomial variables in Eqs. (17), the $m$-dimensional surface contains the straight lines defined by arbitrarily fixing the values of ($m$-1) of the $m$ $p_j$'s in Eqs. (17), while the remaining one is freely varying. Thus, a vertical line appearing at a given frequency in one of the $p_j(\omega)$ representations with the other $p_j$'s having definite values means that the corresponding straight line is fully contained within the intersection with $\{\pm i\omega\}$ and, more specifically, within the ($N$-2)-dimensional subsurface of $\{\pm i\omega\}$ corresponding to the given frequency.

The horizontal line at $p = 1$ aids to distinguish the different classes of fixed points on the represented solutions. As a function of the frequency, the twofold solution describes two classes of fixed points that may be equal or not, and every time one of the $p_j(\omega)$ curves of one of the solutions crosses the $p = 1$ line there is a change of class from $S_j$ to $S_{j\pm1}$ for such a solution. Such crossings correspond to $\{0,\pm i\omega\}$ non-hyperbolic fixed points in the kj space and they may easily occur in the intersection with $\{\pm i\omega\}$ because require a single condition only. Instead the occurrence of $\{\pm i\omega_1, \pm i\omega_2\}$ fixed points may be generically excluded since they would require the twofold condition of identical values of both $p_1$ and $p_2$ at two different frequencies. In the absence of $\{\pm i\omega_1, \pm i\omega_2\}$ fixed points, the continuous frequency intervals associated with a given class of fixed point corresponds to the same bifurcation in the sense of yielding equal dimensional changes in the invariant manifolds of the fixed point. Therefore, to identify the kinds of bifurcations we only need to calculate the fixed point eigenvalues once at each frequency interval of the two solutions. For simplicity we have not included such a kind of information in Fig. B1 but as an illustrative example we say that the third bifurcation of $S_0$, through which it passes from four to six unstable dimensions with increasing the $p_j$ values, occurs in the continuous line solution from the vertical line at $\omega \approx 157$ to the end at $\omega \approx 316$ and in the dotted line solution from the vertical line at $\omega \approx 65$ to the common end at $\omega \approx 316$, with the latter including the imposed bifurcation W7 at $\omega \approx 314$. Since the involved $p_j$ values are moderate and almost all available with $g2$, the family of systems $N6m2$ C $g2$ can sustain such a bifurcation practically over the full range of involved frequencies. In contrast, the $S_2$ class only appears in a narrow interval of the dotted line solution around the imposed bifurcation W4 at w = 3.14, where the fixed point changes from two to four unstable dimensions, and, then, we can exclude the occurrence of its second bifurcation, which is the lacking one in order to exhaust the stable manifolds of the four classes of fixed points. Sometimes the localization of a periodic orbit becomes rather difficult, usually due to the significant influence of a higher frequency mode very close to the fixed point, and it is then convenient to try at different frequencies within the continuous interval (s) of the given bifurcation.

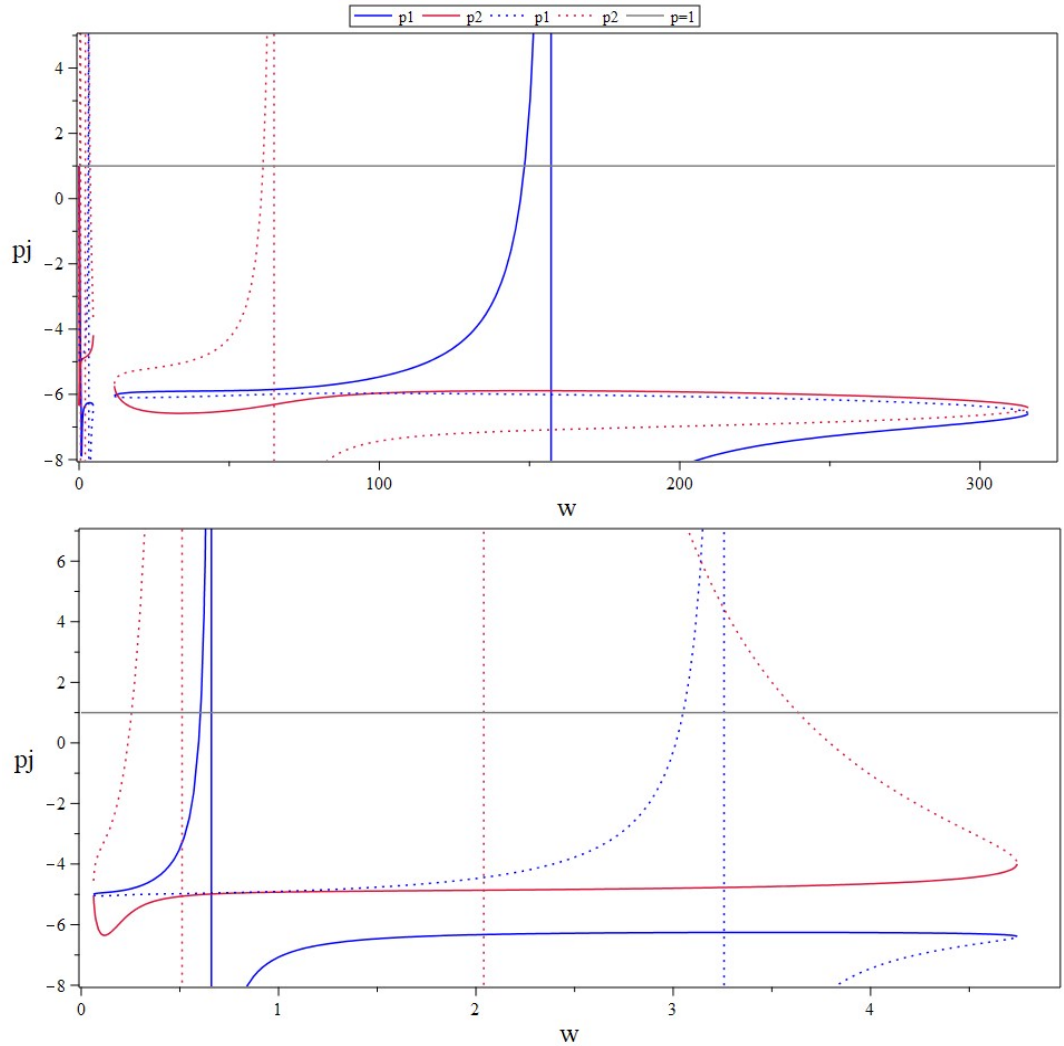

Fig. B1. Representation of the $p_1$ and $p_2$ values of the potential fixed points $\{\pm i\omega\}$ of the system families of case $N6m2$ C, describing the potential Hopf bifurcations of such systems as a function of the bifurcating frequency and containing the seven imposed bifurcations for the design (Table C1). The lower figure is an enlarged detail of the upper one. The continuous and dotted lines distinguish the two different solutions, respectively. The dotted red line includes an inverted U with maximum around $\omega = 1$ and $p_2 = -38$ in between the two vertical lines. The horizontal grey line at $p = 1$ aids to identify the four classes of fixed points.

The fact that the bifurcations may occur within a broad frequency range, accordingly with the corresponding broad ranges of $p_j$ values, means that, in the linear regime, a given periodic orbit may be generated as a harmonic oscillation within a, sometimes very large, frequency range, with the consequent uncertainty about its oscillatory features when going far from the bifurcation by varying the $\beta_j$ values. The nonlinearities will define the actual properties of the periodic orbit as a function of the $\beta_j$ parameters, including both those intrinsic to the given oscillation and those arising from mixing influences of other oscillation modes. In this way, the nonlinearities define the actual features of the oscillation modes of the two-parameter family of systems within the parameter plane and also, since the nonlinear effects depend on the variables, determine for a given system the variability with which the oscillation modes manifest in the different phase space regions where they actuate.

For m = 3, the couple of equations characterizing the intersection with $\{\pm i\omega\}$ may be tackled if, in addition to $\omega$, one of the $p_j$'s, say $p_3$, is taken as a parameter and analytic expressions for $p_1(p_3, \omega)$ and $p_2(p_3, \omega)$ are then obtained that represented as a function of $\omega$ for a given value of $p_3$ provides a twofold solution for $p_1$ and $p_2$ covering continuous frequency intervals, like happens for $m = 2$ (Fig. B1), while in a 3D representation as a function of $\omega$ and $p_3$ the twofold solution covers continuous two-dimensional zones, in correspondence with the fact that in the $k_q$ space the intersection with $\{\pm i\omega\}$ is two-dimensional. By assuming properly designed systems with a significant number of imposed Hopf bifurcations, we expect that the described behaviour of the Hopf bifurcation will extend to $m$-parameter families of systems with higher $m$ values.

# Appendix C

## Designed Systems for the Simulations and Technical Design Details

The Tables C1 and C2 characterize the systems used for the reported simulations. Table C1 describes the set of imposed Hopf bifurcations to design the systems and includes the additional bifurcations that the designed systems can sustain in accordance with our purpose of exhausting the stable manifold. Table C2 gives the coefficients $c_{j,q}$ and $d_{j,q}$ of the different systems distinguishing those obtained through the design process from those whose value has been predefined, either by directly choosing it or through the assumed condition $d_{j,N-m+j} = -c_{j,N-m+j}$. While, for $m = 1$, the $c_{1,q}$ negative signs and the $d_{1,q}$ sign alternation reflects the well-organized structure of competing effects underlying the proper oscillatory scenario unfolding, for $m = 2$, we have been unable to appreciate any distinguishing criterion between proper and improper designed systems through their coefficients, and this refers to Table C2 and to a large number of other designed systems. More apparent is the influence of the distribution of imposed bifurcations among the different classes of fixed points. To illustrate this point with concreteness, let us consider the case of $N6m2$ systems potentially possessing four different classes of fixed points: $S_0$, $S_{1a}$, $S_{1b}$ and $S_2$, upon which up to nine Hopf bifurcations could be imposed up to exhaust their stable manifolds: three on $S_0$ and two on the other three classes. By considering the bifurcations ordered according to their frequency $\omega_j$, from lower to higher, we can impose the nine successive bifurcations, for instance, on $S_0$, $S_{1a}$, $S_{1b}$, $S_2$, $S_0$, $S_{1a}$, $S_{1b}$, $S_2$ and $S_0$, by appropriately choosing the corresponding $(p_1,p_2)$ values within the corresponding classes, but just with such values we can adjust the bifurcations order on each fixed point in relation to how they destabilize successive pairs of dimensions, independently of the frequency order. For simplicity we have usually imposed the successive bifurcations on a fixed point ordered according to their frequency from lower to higher and this is the case for all the systems of Table C2, but we have verified that properly designed systems are also obtained by inverting the frequency order of the imposed bifurcations in one or more fixed points. On the other hand, the number of free coefficients in the system (5) of differential equations allows for eight bifurcations only and, as it will be discussed in more detail bellow, our capability in polynomial system solving for $N6m2$ systems appears limited to fourteen bifurcations. Thus, the $N6m2$ systems C, D and E have been designed by imposing seven bifurcations and by arbitrarily choosing the values of two arbitrarily chosen $c_{j,q}$ coefficients in addition to $c_{11}$ and $c_{22}$. For D and E the bifurcations distribution among fixed point classes was that mentioned above while for C the seventh bifurcation was imposed on $S_0$ instead of $S_{1b}$. The sets of imposed $(p_1,p_2)$ values have been taken equal for the different systems in order to facilitate the dynamical analysis of successive systems. Notice the symmetric choice of the $(p_1,p_2)$ values for the two $S_1$ classes and also that, in D and E, the symmetry extends to the bifurcation frequencies, while in C the frequencies are different. The frequency symmetry between $S_{1a}$ and $S_{1b}$ is also imposed in the system $N5m2$ B. Not always but it is usual that the defined systems exhibit additional bifurcations contributing to exhaust the stable manifolds of the fixed points and this is the case for all the $m = 2$ systems of Table C1 (The reported values have been chosen at convenience within the continuous interval of each extra bifurcation). In this way, systems B, D and E may exhaust fully the stable manifolds of the four classes of fixed point, although in the three cases the last bifurcation requires an enormous $p_j$ value so that it will be irrelevant in practice. Instead the additional bifurcation at W8 in systems C, D, and E corresponds to moderate enough $p_j$ values and it may then significantly influence the dynamical behaviour of the system family, as it is the case for the reported examples with the nonlinear function $g2$. The ninth and pending bifurcation for system C should correspond to $S_2$ and, as discussed in relation to Fig. B1, it cannot occur within the two-parameter plane. Oddly enough, the dynamics of C reported in Figs. 8-10 exhibits a significant contribution of a ninth mode whereas it is absent in D and E.

Table C1. Chosen values of ($p_1$, .., $p_m$) and $w$ for the imposed Hopf bifurcations in the design of the systems used in the simulations. Wj denotes periodic orbits. In blue the corresponding class of fixed point. In red the extra bifurcations.

Table C2. Coefficients $c_{j,q}$ and $d_{j,q}$ of the designed systems used in the simulations. These values have been obtained by rounding the calculated ones (to six digits for $m = 2$). In red the non-free coefficients. The number of free coefficients is twice that of imposed bifurcations (see Table C1) and which for $m = 2$ is usually lower than the maximum allowable in the four classes of fixed points, being limited by the polynomial system solver capability.

### N5m1 A

| Wj | | $p_1$ | $w$ | $2\pi/w$ |
|---|---|---|---|---|
| 1 | $S_0$ | -5 | 0.05 | 125.7 |
| 2 | $S_1$ | +6 | 0.3 | 20.9 |
| 3 | $S_0$ | -6.4 | 1.8 | 3.5 |
| 4 | $S_1$ | +7 | 26 | 0.24 |

### N5m2 B

| Wj | | $p_1$ | $p_2$ | $w$ | $2\pi/w$ |
|---|---|---|---|---|---|
| 1 | $S_0$ | -5 | -5 | 0.0628 | 100 |
| 2 | $S_{1a}$ | -5 | +6 | 0.314 | 20 |
| 3 | $S_{1b}$ | +6 | -5 | 0.314 | 20 |
| 4 | $S_2$ | +6 | +6 | 1.26 | 5 |
| 5 | $S_0$ | -6 | -6 | 6.28 | 1 |
| 6 | $S_{1b}$ | +7.85 | -6.85 | 47.5 | 0.13 |
| 7 | $S_{1a}$ | -241 | +6.52 | 322 | 0.019 |

### N6m2 C

| Wj | | $p_1$ | $p_2$ | $w$ | $2\pi/w$ |
|---|---|---|---|---|---|
| 1 | $S_0$ | -5 | -5 | 0.0628 | 100 |
| 2 | $S_{1a}$ | -5 | +6 | 0.314 | 20 |
| 3 | $S_{1b}$ | +6 | -5 | 0.628 | 10 |
| 4 | $S_2$ | +6 | +6 | 3.14 | 2 |
| 5 | $S_0$ | -6 | -6 | 12.57 | 0.5 |
| 6 | $S_{1a}$ | -6 | +6.5 | 62.83 | 0.1 |
| 7 | $S_0$ | -6.5 | -6.5 | 314 | 0.02 |
| 8 | $S_{1b}$ | +4.29 | -5.89 | 152 | 0.04 |

### N6m2 D

| Wj | | $p_1$ | $p_2$ | $w$ | $2\pi/w$ |
|---|---|---|---|---|---|
| 1 | $S_0$ | -5 | -5 | 0.0628 | 100 |
| 2 | $S_{1a}$ | -5 | +6 | 0.314 | 20 |
| 3 | $S_{1b}$ | +6 | -5 | 0.314 | 20 |
| 4 | $S_2$ | +6 | +6 | 1.57 | 4 |
| 5 | $S_0$ | -6 | -6 | 12.57 | 0.5 |
| 6 | $S_{1a}$ | -6 | +6.5 | 314 | 0.02 |
| 7 | $S_{1b}$ | +6.5 | -6.5 | 314 | 0.02 |
| 8 | $S_2$ | +7.3 | +7.3 | 4328 | 0.0014 |
| 9 | $S_0$ | -280 | -6.46 | 3646 | 0.0017 |

### N6m2 E

| Wj | | $p_1$ | $p_2$ | $w$ | $2\pi/w$ |
|---|---|---|---|---|---|
| 1 | $S_0$ | -5 | -5 | 0.0628 | 100 |
| 2 | $S_{1a}$ | -5 | +6 | 0.628 | 10 |
| 3 | $S_{1b}$ | +6 | -5 | 0.628 | 10 |
| 4 | $S_2$ | +6 | +6 | 3.14 | 2 |
| 5 | $S_0$ | -6 | -6 | 31.4 | 0.2 |
| 6 | $S_{1a}$ | -6 | +6.5 | 314 | 0.02 |
| 7 | $S_{1b}$ | +6.5 | -6.5 | 314 | 0.02 |
| 8 | $S_2$ | +8.46 | +8.46 | 4092 | 0.0015 |
| 9 | $S_0$ | -120 | -6.79 | 2588 | 0.0024 |

### N5m1 A

| $q$ | $c_{1,q}$ | $d_{1,q}$ |
|---|---|---|
| 1 | -20 | 2.9 |
| 2 | -323 | -50 |
| 3 | -59 | 10 |
| 4 | -26 | -5.1 |
| 5 | -0.046 | 0.046 |

### N5m2 B

| $q$ | $c_{1,q}$ | $c_{2,q}$ | $d_{1,q}$ | $d_{2,q}$ |
|---|---|---|---|---|
| 1 | -500 | 360.895 | -8.36751 | 112.813 |
| 2 | 688.537 | -500 | 11.9384 | -155.551 |
| 3 | 10 | -10 | 2.10513 | -2.8886 |
| 4 | 2.85421 | 0 | -2.85421 | 0 |
| 5 | 0 | -0.0206974 | 0 | 0.0206974 |

### N6m2 C

| $q$ | $c_{1,q}$ | $c_{2,q}$ | $d_{1,q}$ | $d_{2,q}$ |
|---|---|---|---|---|
| 1 | -200 | 3000 | -33.3135 | 509.704 |
| 2 | -3000 | -200 | -493.788 | -28.9993 |
| 3 | 707.506 | 90.92 | 141.126 | 38.5204 |
| 4 | -7786.2 | 103502 | -1257.74 | 21003.3 |
| 5 | 0.0975128 | 0 | -0.0975128 | 0 |
| 6 | 0 | 386.052 | 0 | -386.052 |

### N6m2 D

| $q$ | $c_{1,q}$ | $c_{2,q}$ | $d_{1,q}$ | $d_{2,q}$ |
|---|---|---|---|---|
| 1 | -200 | -3000 | -3.23244 | 214.508 |
| 2 | -3000 | -200 | -491.81 | 58.1074 |
| 3 | -12849700 | -850748 | -2106690 | 247847 |
| 4 | 10.8482 | 0.371201 | 10.1695 | -0.438213 |
| 5 | -34726.9 | 0 | 34726.9 | 0 |
| 6 | 0 | -0.0400246 | 0 | 0.0400246 |

### N6m2 E

| $q$ | $c_{1,q}$ | $c_{2,q}$ | $d_{1,q}$ | $d_{2,q}$ |
|---|---|---|---|---|
| 1 | -152 | 200 | -5.58776 | 2429.77 |
| 2 | -200 | -152 | -32.7155 | 41.6331 |
| 3 | -2886640 | -2125490 | -472240 | 589097 |
| 4 | 24.822 | 6.61275 | 3.34745 | -6.15871 |
| 5 | -7582.04 | 0 | 7582.04 | 0 |
| 6 | 0 | -0.545635 | 0 | 0.545635 |

The systems of polynomial equations have been tried to be solved with Maple, usually numerically with the *fsolve* command since it can deal with more equations than the algebraic solver *RootFinding* [*Isolate*]. The *fsolve* command outputs a single real root (a few may be achieved by using the *avoid* option) but we don't need to find all the solutions since one or, better, a few of them are enough for our purpose of designing dynamical systems. Dealing with the system of dynamical equations (5), the polynomial equations are of degree $m$. For $m$ = 2 the maximum number of equations for which *fsolve* provide us with solutions depends on the dynamical dimension: up to fourteen for $N = 6$ and up to eighteen for $N = 8$, while with *Isolate* we don't surpass ten equations although obtaining around 25 solutions. The computations have been done with 20 or 25 digits to assure enough precision in obtaining the roots of the polynomials but such a precision is unnecessary for the good working of the designed dynamical systems and the obtained coefficients may be drastically rounded without significant changes in the dynamical behaviour. Concretely, the reported simulations for $m$ = 2 have been done with coefficients rounded to six digits without noticeable variations in the frequencies and $p_j$ values of the Hopf bifurcations with respect to those imposed in the design process.

Of course the design success depends also on the chosen values for the predefined coefficients and this opens a wide way of possibilities. Our trials with $m$ = 2 have been usually done by defining the value of some $c_{j,q}$ coefficients in addition to the $c_{11}$ and $c_{22}$: two for $N = 6$ and six for $N = 8$. A few trials with $d_{j,q}$ coefficients pointed out that their values are more critical for the design success.